\documentclass[usenatbib,usegraphicx]{mn2e}
\usepackage{times}
\usepackage{graphicx, epsfig}
\usepackage[fleqn]{amsmath}
\usepackage{amsfonts}
\usepackage{amssymb}
\usepackage[totalwidth=17.8cm, totalheight=24.0cm]{geometry}

\title[Kinemetry]{Kinemetry: a generalisation of photometry to the
  higher moments of the line-of-sight velocity distribution}

\author[Davor Krajnovi\'c et al.]
       {Davor Krajnovi\'c,$^{1,2}$\thanks{E-mail: dxk@astro.ox.ac.uk}
       Michele Cappellari,$^{1}$
       P. Tim de Zeeuw,$^{1}$ 
       Yannick Copin$^{3}$\\
$^1$Sterrewacht Leiden, Postbus 9513, 2300 RA Leiden, The Netherlands\\
$^2$Denys Wilkinson Building, University of Oxford, Keble Road, OX1 3RH, UK \\
$^3$Institut de Physique Nucl\'eaire de Lyon, 69622, Villeurbanne, France}

\pagerange{\pageref{firstpage}--\pageref{lastpage}} \pubyear{2005}

\newcommand{\SAURON}{{\tt SAURON} }

\newcommand{\kms}{\>{\rm km}\,{\rm s}^{-1}}

\newcommand{\ud}{\mathrm{d}}

\def\aj{AJ}             % Astronomical Journal
\def\apj{ApJ}           % Astrophysical Journal
\def\apjl{ApJ}          % Astrophysical Journal, Letters
\def\apjs{ApJS}         % Astrophysical Journal, Supplement
\def\aap{A\&A}          % Astronomy and Astrophysics
\def\aaps{A\&AS}        % Astronomy and Astrophysics, Supplement
\def\mnras{MNRAS}       % Monthly Notices of the RAS
\begin{document}
\label{firstpage}
\maketitle

\begin{abstract}
  We present a generalisation of surface photometry to the
  higher-order moments of the line-of-sight velocity distribution of
  galaxies observed with integral-field spectrographs. The
  generalisation follows the approach of surface photometry by
  determining the best fitting ellipses along which the profiles of
  the moments can be extracted and analysed by means of harmonic
  expansion. The assumption for the odd moments (e.g. mean velocity)
  is that the profile along an ellipse satisfies a simple cosine law.
  The assumption for the even moments (e.g velocity dispersion) is
  that the profile is constant, as it is used in surface photometry.
  We test the method on a number of model maps and discuss the meaning
  of the resulting harmonic terms. We apply the method to the
  kinematic moments of an axisymmetric model elliptical galaxy and
  probe the influence of noise on the harmonic terms. We also apply
  the method to {\tt SAURON} observations of NGC~2549, NGC~2974,
  NGC~4459 and NGC~4473 where we detect multiple co- and
  counter-rotating (NGC~2549 and NGC~4473 respectively) components. We
  find that velocity profiles extracted along ellipses of early-type
  galaxies are well represented by the simple cosine law (with 2\%
  accuracy), while possible deviations are carried in the fifth
  harmonic term which is sensitive to the existence of multiple
  kinematic components, and has some analogy to the shape parameter of
  photometry. We compare the properties of the kinematic and
  photometric ellipses and find that they are often very similar, but
  a study on a larger sample is necessary.  Finally, we offer a
  characterisation of the main velocity structures based only on the
  kinemetric parameters which can be used to quantify the features in
  velocity maps.
\end{abstract}

%________________________________________________________________

%%%%%%%%%%%%%%%%%%%%%%%%%%%%%%%%%%%%%%%%%%%%%%%%%%%%%%%%%%%%%%%%%%%%
%
%SECTION SECTION1 SECTION1 SECTION1 SECTION1 SECTION1 SECTION1 
%
%%%%%%%%%%%%%%%%%%%%%%%%%%%%%%%%%%%%%%%%%%%%%%%%%%%%%%%%%%%%%%%%%%%%

\section{Introduction}
\label{s:intro}

Over the last three decades broad band observations of early-type
galaxies were successfully analysed by a method commonly called
surface photometry or, simply, photometry. This method is based on the
analysis of isophotal shapes of the projected surface brightness. The
development of the method was stimulated by the empirical discovery
that the isophotes of early-type galaxies are reproduced by ellipses
to better than 1 per cent \citep{1984ApJS...56..105K,
  1985ApJS...57..473L, 1985AJ.....90..169D, 1987MNRAS.226..747J,
  1990AJ....100.1091P}.  Although the isophotes are elliptical in
shape to high accuracy, under careful examination, many isophotes of
early-type galaxies do show differences from pure ellipses at a level
of $\approx0.5\%$ \citep[e.g.][]{1988A&AS...74..385B,
  1990AJ....100.1091P}. The true success of photometry was the ability
to measure these deviations and classify early-type galaxies
accordingly into disky and boxy objects
\citep{1985MNRAS.216..429L,1987MNRAS.226..747J,1987A&A...177...71B}.
When combined with information on total luminosity and spatially
resolved spectroscopy, it followed that the duality of photometric
properties of early-type galaxies is reflected in a duality of
kinematic properties, where faint disky objects were found to rotate
faster than luminous boxy objects \citep{1983ApJ...266...41D,
  1983ApJ...266..516D, 1988A&A...193L...7B, 1988A&A...195L...1N,
  1988A&A...195L...5W, 1989A&A...217...35B, 1989A&A...215..266N,
  1992A&A...262...52B, 1994MNRAS.269..785B}. High resolution imaging
studies with the Hubble Space Telescope \citep{1994AJ....108.1567J,
  1995AJ....110.2622L, 2001AJ....121.2431R}, again based on
photometric analysis, revealed new properties of early-type galaxies
that deepened the division of the galaxies in two groups. These
remarkable set of discoveries resulted in a revised classification
scheme of galaxies \citep{1996ApJ...464L.119K}.

It is fair to say that our increased knowledge of early-type galaxies
(partially) comes from photometry and the ability of the method to
harvest and describe in a compact way the information from
two-dimensional images. However, the integrated light remains a
limited source of information about the internal structure and thus
the true nature of galaxies. Kinematic information is also essential
to fathom the complexity of galaxies and, especially, two-dimensional
kinematic information is required \citep[e.g.][]{1991ApJ...383..112F,
  1991AJ....102..882S, 1994ApJ...425..481S, 1994ApJ...425..458S,
  1994ApJ...425..500S, 1994MNRAS.271..924A}.

Two-dimensional velocity maps were, until recently, only possible for
objects with clear emission lines. Such maps were the result of
studies of, e.g., H{\small I} with radio interferometers,
\citep[e.g.][]{2002A&A...390..829S}, CO with millimetre
interferometers \citep[e.g.][]{2003ApJS..145..259H} and H$\alpha$ with
Fabry-Perot spectrographs \citep[e.g.][]{2005MNRAS.360.1201H}.  The
advent of integral-field spectrographs (e.g. {\tt TIGER}, {\tt OASIS},
{\tt SAURON}, {\tt PMAS}, {\tt GMOS}, {\tt SINFONI}, {\tt OSIRIS}),
however, has brought two-dimensional kinematic measurements to
classical optical and near-infrared wavelengths. Here it is possible
to probe both the stellar absorption- and gas emission-lines, which
may co-exist in the same potential with very different spatial
distributions and dynamical structures. The wealth of features seen in
stellar kinematic maps of early-type galaxies
\citep{2004MNRAS.352..721E} confirms the usefulness of two-dimensional
data, but also poses a problem to efficiently harvest and interpret
the important features from the maps.

An approach using harmonic expansion was developed for the analysis of
two-dimensional velocity maps of disk galaxies. This method divides a
velocity map into individual rings \citep[the so-called tilted-ring
method,][]{1987PhDT.......199B} and performs a harmonic expansion
along these rings \citep[e.g.][]{1978MNRAS.183..779B,
  1991wdir.conf...40T,1994ApJ...436..642F,
  1997MNRAS.292..349S,2004ApJ...605..183W}. This analysis, however, is
based on the assumption that the emission-line emitting material is
confined to a thin disk structure.  Clearly, a spheroidal distribution
of stars typical of early-type galaxies does not have the same
dynamical properties as a gas disk. To explore those intrinsic
properties, one needs a more general method which will not be based on
assumptions about the nature of the observed system (e.g. a disk), but
will rely solely on the properties of the investigated observables
(e.g. surface brightness or velocity).

Here we explore such a general method.  Photometry earned its spurs in
intensive applications over the last three decades and offers a
natural starting point for the analysis of two-dimensional kinematic
maps, although one can, in principle, invent a number of basis
functions which can describe a two-dimensional distribution. In this
paper, we present a generalisation of photometry to the higher-order
moments of the line-of-sight velocity distribution (LOSVD). This
generalisation is based on the theoretical fact that surface
brightness is itself a moment of the LOSVD, and on our empirical
discovery that the velocity maps of many early-type galaxies can be
well reproduced by a simple cosine law along sampling ellipses.  We
call our method {\it kinemetry}\footnote{This name was introduced by
  \citet{2001sf2a.conf..289C} who presented a preliminary discussion
  on this topic.}, which reduces to photometry for the investigation
of surface brightness distributions.

In Section~\ref{s:back} we present the theoretical background and
motivation for the new method. Section~\ref{s:method} presents the
technical aspects of the method. The meaning of the kinemetric
coefficients and their diagnostic merits for different model maps are
presented in Section~\ref{s:kinpar}. In Section~\ref{s:appli} we apply
the method on kinematic maps of an axisymmetric model galaxy, present
the application of kinemetry to actual observations and characterise
typical structures on velocity maps.  We summarise our conclusions in
Section~\ref{s:sum}.

%%%%%%%%%%%%%%%%%%%%%%%%%%%%%%%%%%%%%%%%%%%%%%%%%%%%%%%%%%%%%%%%%%%%
%
%SECTION2 SECTION2 SECTION2 SECTION2 SECTION2 SECTION2 SECTION2 
%
%%%%%%%%%%%%%%%%%%%%%%%%%%%%%%%%%%%%%%%%%%%%%%%%%%%%%%%%%%%%%%%%%%%%

\section{Theoretical background and motivation}
\label{s:back}
The dynamics of a collisionless stellar system is fully specified by
its phase-space density or distribution function $f = f(\vec{x},
\vec{v},t)$ \citep[e.g.][]{1987gady.book.....B}, but this quantity is
not measurable directly. When observing external galaxies, we measure
properties that are integrated along the line-of-sight (LOS).  An
additional complication is that the galaxies are viewed from a certain
direction, and we actually observe only those projected properties of
the integrated distribution function.

The observables, which reveal only averages over a large number of
unresolved stars, are the surface brightness and the LOSVD, sometimes
also called the velocity profile\footnote{In this work, however, the
  term `velocity profile' (or generally, kinematic profile) will be
  used to denote a set of velocity (kinematic) measurements extracted
  from a map along a certain curve.} \citep[for a review
see][]{1994feg..conf..231D}. The surface brightness is given by
\begin{equation}
  \label{eq:surf}
 \mu(x,y) = \int_{\mathrm{LOS}}
\ud z \int\! \int\! \int\! \ud \vec{v}\,f(\vec{r},\vec{v}).
\end{equation}
The rest of the observable information is carried by the LOSVD, which
relates to the distribution function via
\begin{equation}
  \label{eq:losvd}
  \mathcal{L}(v;x,y) = \int_{\mathrm{LOS}} \ud z
                                 \int\! \int\! \ud v_{x} \ud v_{y}\,f(\vec{r},\vec{v}), 
\end{equation}
\noindent where (\emph{x,y,z}) are the three spatial coordinates,
oriented such that the LOS is along the \emph{z}-axis. From these
equations it is trivial to show that surface brightness is the zero-th
order moment of the LOSVD. Higher-order moments, such as $\langle V
\rangle$, or $\langle V^2 \rangle$, can easily be related to the
observables of the LOSVD: mean velocity $V$, velocity dispersion
$\sigma$ and higher-order moments, commonly parametrised by
Gauss-Hermite coefficients \citep{1993ApJ...407..525V,
  1993MNRAS.265..213G}, $h_{3}$ and $h_{4}$ being the most used.

The kinematic moments of stationary triaxial systems show a high
degree of symmetry which can be expressed through their parity. The
mean velocity is an odd moment, while the velocity dispersion is an
even moment.  In practice, this means that a two-dimensional map of a
given moment shows corresponding symmetry. Maps of even moments are
\emph{point-symmetric}, while maps of odd moments are
\emph{point-anti-symmetric}. In polar coordinates on the sky plane
this gives:
\begin{eqnarray}
  \label{eq:paspolar}
  \mu_{e}(r,\theta+\pi) & = & \mu_{e}(r,\theta)\nonumber,\\
  \mu_{o}(r,\theta+\pi) & = &-\mu_{o}(r,\theta),
\end{eqnarray}

\noindent where $\mu_{e}$ and $\mu_{o}$ are arbitrary even and odd
moments of the LOSVD, respectively. Furthermore, if the observed
system is axisymmetric, the even moment of the LOSVD will also be
\emph{mirror-symmetric} or, correspondingly, an odd moment will be
\emph{mirror-anti-symmetric}:
\begin{eqnarray}
  \label{eq:maspolar}
  \mu_{e}(r,\pi - \theta) & = & \mu_{e}(r,\theta)\nonumber,\\
  \mu_{o}(r,\pi - \theta) &=& -\mu_{o}(r,\theta).
\end{eqnarray}

\noindent
A kinematic moment that satisfy both symmetries
(eqs.~\ref{eq:paspolar} and~\ref{eq:maspolar}) is said to be
bi-(anti)-symmetric. The existence of these symmetries can be used to
simplify the harmonic analysis. Point-symmetric moments give rise only
to even harmonic terms, while point-anti-symmetric moments will
require only odd harmonic terms. Mirror-(anti)-symmetry additionally
requires that there is no change in the moment's position angle. For a
more detailed discussion on the influence of these equations on the
terms of the harmonic expansion see Appendix~\ref{s:circ}.

Following these conventions, surface brightness, as the zeroth order
moment, is an even moment. As the surface brightness and the kinematic
moments of the LOSVD are moments of the same distribution function, it
is natural to analyse them similarly, and we can ask in which way one
can generalise the method of photometry to work on the higher moments
of the LOSVD. Considering symmetry, the even moments do not require
much change in the method, but the odd moments need a new working
assumption. Also, while the isophotes of integrated light are
elliptical to a high accuracy, the choice of sampling curves is not as
obvious for odd kinematic moments.

Since the light distribution of early-type galaxies is elliptical it
is reasonable to assume that an expansion along ellipses would also be
suitable for the extraction of profiles from kinematic moment maps
giving some insight in their structure and properties. The choice of
the actual ellipticity, however, is somewhat arbitrary. For example,
it is possible to expand along the galaxy isophotes or along ellipses
that correspond to deprojected circles.  In the first case, one
follows the photometry and probes regions of the same projected
surface brightness. In the second case, one samples locations with
equal intrinsic radii, but it is necessary to assume an inclination
for the galaxy, which is usually difficult to estimate. It may also
not be physically justified to use a constant ellipticity for the
whole velocity map, as the ellipticity of the light distribution may
change with radius.

As a special case, it is possible to expand a kinematic map along
circles. This approach was investigated by
\citet{2001sf2a.conf..289C}, \citet{2002ASPC..282..508C} and
\citet{2004KrajnovicThesis}, and although it is straight-forward and
requires no a priori assumption, a large number of harmonic terms are
often necessary, preventing a simple description of the maps and
complicating the analysis (see Appendix~\ref{s:circ} for a
discussion).  Additionally, an expansion along ellipses can reduce to
expansion along circles in limiting cases (see
section~\ref{ss:special}).  \citet{1989AJ.....98..538F} also note the
advantages of extraction along circles, even for photometry, but
conclude that an approach based on ellipses is superior.

Consider the first moment of the LOSVD, the mean velocity.  In
specific cases, the geometry of the system offers a choice for curves
along which velocity profiles can be extracted. If a velocity map
traces material in a thin disk, such as is often the case in H{\small
  I}, CO and H$\alpha$ gas emission maps, velocity can be well
represented by circular motion.  If the inclination of the disk is
known, a circular orbit in the plane of the galaxy can be projected as
an ellipse on the sky. A velocity profile extracted along this ellipse
is then of simple cosine form, which can be expressed by
\begin{equation}
  \label{eq:tilt}
  V(R,\psi)= V_{0} + V_{C}(R) \sin i \cos \psi,
\end{equation}

\noindent where $R$ is the radius of a circular ring in the plane of the
galaxy (or the semi-major axis length of the ellipse on the sky),
$V_{0}$ is the systemic velocity, $V_{C}$ is the ring circular
velocity, $i$ is the ring inclination ($i=0$ for a ring seen face-on)
and $\psi$ is the azimuthal angle measured from the projected major
axis in the plane of the galaxy. The inclination $i$ is related to the
axial ratio or flattening, $q$, of the ring's major and minor axes.
For a given flattening, the ring's velocity profile is most similar to
the velocity of a circular orbit with radius $R$ and inclination $\cos
i = q$. In other words, the flattening defines an ellipse on the sky,
with ellipticity $\epsilon = 1 - q$, which corresponds to a circle
describing the orbits in the plane of the galaxy.

If one assumes that the velocity profiles extracted from velocity maps
(either stellar or gaseous) of early-type galaxies can be described by
the simple cosine law of eq.~(\ref{eq:tilt}) to a high accuracy, then
the first step in the analysis of velocity maps is to determine those
ellipses along which the velocity profile is best described by
eq.~(\ref{eq:tilt}). In analogy with photometry, the next step would
be to determine the deviations of the velocity profile from the cosine
law, which can be achieved by the higher-order Fourier analysis.

We discovered that the velocity profiles extracted from the maps of
the {\tt SAURON} sample can indeed be well described by the cosine law
of eq.~(\ref{eq:tilt}). Note that this observation suggests that the
velocity maps of early type galaxies closely resemble the observed
kinematics of inclined circular disks. One has, however, to take a
step back and recognise that spheroids are not disks, and the two
cases should be clearly differentiated during the interpretation of
the results.  The fact that the velocity maps of galaxies are well
approximated by the thin disk model allows us to develop a simple
generalisation of photometry based on eq.~(\ref{eq:tilt}), but also
reflects an interesting property of the internal structure of
galaxies, which is worth exploring further (see
section~\ref{ss:ngc4473} for a few examples while the results for the
rest of the {\tt SAURON} sample will be reported in a future paper).

The method presented in this paper is applicable to all moments of the
LOSVD. The quality of kinematic data, however, usually decreases
rapidly with increasing order of the kinematic moment: the signal is
generally simply too weak for a good enough extraction of the
higher-order moments from raw data. Perhaps it is even fair to say
that only the mean velocity maps reach the quality of photometric
observations from thirty or so years ago, when surface photometry was
introduced.  For these reasons, as well as for presentation purposes,
we restrict our analysis to the mean velocity maps. The method,
however, can be straight-forwardly applied to the higher order moments
(see Section~\ref{ss:jeans} for an example).

%%%%%%%%%%%%%%%%%%%%%%%%%%%%%%%%%%%%%%%%%%%%%%%%%%%%%%%%%%%%%%%%%%%%%
%
%SECTION3 SECTION3 SECTION3 SECTION3 SECTION3 SECTION3 SECTION3 
%
%%%%%%%%%%%%%%%%%%%%%%%%%%%%%%%%%%%%%%%%%%%%%%%%%%%%%%%%%%%%%%%%%%%%%

\section{The method}
\label{s:method}

In this section we present the details of the new algorithm extending
surface photometry to the maps of odd moments of the LOSVD.

\subsection{Harmonic expansion}
\label{ss:exp}

Fourier analysis is the most straight-forward approach to characterise
any periodic phenomenon. A velocity map $K (a, \psi)$ can be divided
into a number of elliptical rings yielding velocity profiles which can
be described by a finite (and small) number ($N+1$) of harmonic terms
(frequencies):
\begin{equation}
  \label{eq:kinlin}
  K(a,\psi) = A_{0}(a) + \sum_{n=1}^{N} A_{n}(a)\,\sin(n \psi) +
                                B_{n}(a)\,\cos(n \psi),
\end{equation}

\noindent where $\psi$ is the eccentric anomaly, that in case of disks
corresponds to the azimuthal angle of eq.~(\ref{eq:tilt}), and $a$ is
the length of the semi-major axis of the elliptical ring.

As presented by \citet{1978MNRAS.182..797C},
\citet{1983ApJ...266..562K} and \citet{1987MNRAS.226..747J}, to find
the best sampling ellipse along which a profile $\mu_{e}$ should be
extracted, one minimises the harmonic terms sensitive to the
parameters of the ellipse (centre, position angle and ellipticity for
a given semi-major axis length). In the case of an even moment (e.g.
surface brightness) the best fitting ellipse parameters are given by
minimising $A_1$, $B_1$, $A_2$ and $B_2$ coefficients of the harmonic
expansion. This means that a photometric ellipse is determined when a
sufficiently good fit to $\mu_{e}(a, \psi)= A_{0}(a)$ is obtained. In
the case of an odd moment of the LOSVD (e.g. mean velocity), given our
empirical finding, the sampling ellipse parameters are determined by
requiring that the profile along the ellipse is well described by
eq.~(\ref{eq:tilt}), or, generally, $\mu_{o}(a, \psi)= B_1(a) \cos
\psi$.

Ellipse parameters (position angle, centre and flattening) can be
determined by a minimising a small number of harmonic terms.  In
Appendix~\ref{s:influ} we investigate which terms are sensitive to
changes of the ellipse parameters. We find that:

\begin{itemize}
\item[-] an incorrect kinematic centre produces non-zero $A_0$, $A_2$,
  $B_2$, and to a lower level $A_1$, $A_3$ and $B_3$ coefficients,
\item[-] an incorrect flattening (inclination) produces a non-zero
  $B_3$ coefficient,
\item[-] an incorrect position angle produces non-zero $A_1$, $A_3$
  and $B_3$ coefficients.
\end{itemize}

\noindent

In the study of disk velocity maps, \citet{1997MNRAS.292..349S} in
their Appendix A2, discuss the same sensitivity, but for small
deviations. In this limit, our results reduce to the analytically
derived conclusions of \citet{1997MNRAS.292..349S}. It is, however,
important to note, that whatever the combination of incorrect centre,
flattening or position angle, only the $n \le 3$ harmonic terms are
affected.

\subsection{The algorithm}
\label{ss:alg}

Given our discovery that the cosine law along sampling ellipses
describes well velocity maps of galaxies, a small number of harmonic
terms is needed to determine the best fitting ellipse. In this case
eq.~(\ref{eq:kinlin}) reduces to
\begin{eqnarray}
  \label{eq:kinNew}
  K(\psi) = A_{0} + A_{1}\,\sin(\psi) + B_{1}\,\cos(\psi)\nonumber\\
               + A_{2}\,\sin(2\psi) + B_{2}\,\cos(2\psi)\nonumber\\
               + A_{3}\,\sin(3\psi) + B_{3}\,\cos(3\psi) 
\end{eqnarray}
\noindent 
where $\psi$ is the eccentric anomaly. In practice, as it is done in
photometry, we sample the points uniformly in $\psi$ and the ellipse
coordinates are then given by $x_e = a \cos\psi$ and $y_e = q\, a
\sin \psi$. This implies that, if the ellipse is projected onto a
circle, the sampling points are equidistant. The selection of
$K(\psi)$ values is obtained by a bilinear interpolation of the
observed map, also in the same way it is usually done in photometry
\citep{1987MNRAS.226..747J}.

The parameters of the best sampling ellipse for an odd kinematic
moment are obtained by minimising:
\begin{equation}
  \label{eq:cchi2}
  \chi^2 = A^2_1 + A^2_2 + B^2_2 + A^2_3 + B^2_3
\end{equation}
from eq.~(\ref{eq:kinNew}).  This non-linear fit is performed in two
stages. The first stage consists of minimising $\chi^2$ on an input
grid of flattenings $q$ and position angles $\Gamma$. The second stage
repeats the fit using a non-linear least-square curve fitting
algorithm, but now also fitting for the coordinates of the centre,
taking as initial values $q=q_{min}$ and $\Gamma=\Gamma_{min}$ from
the first stage and an estimate of the centre ($x_0,y_0$). In this
way, the true $\chi^2$ minimum can be robustly determined and, hence,
the parameters of the best sampling ellipse: centre ($x_c, y_c$),
flattening (ellipticity) and position angle. We use the MINPACK
implementation \citep{1980ANL.....80.74M} of the Levenberg-Marquardt
method\footnote{We used an IDL version of the code written by Craig B.
  Markwardt and available from:
  http://astrog.physics.wisc.edu/$\sim$craigm/idl}. This two stage
approach was implemented to improve the robustness of the method,
preventing the minimisation program from choosing possible secondary
minima.

The choice for the length of semi-major axis of consecutive ellipses
is set by the decrease of surface brightness of the observed object.
The kinematic maps are necessarily spatially binned (e.g. using
adaptive Voronoi binning \citep{2003MNRAS.342..345C}, as in the
examples of Section~\ref{s:appli}), and as the bins typically increase
in size with radius, the distance between the rings has to be
increased to avoid correlation between the ring parameters. The
correlation should also be avoided in the centre where the bins are
usually identical to the original pixels. For these reasons, we
adopted a combination of linear and logarithmic increase of the
semi-major axis length given by this expression: $a = n+1.1^n$, where
$n=1,2,3...$.

The final ellipse parameters obtained by minimisation are then used to
describe an elliptical ring from which a kinematic profile is
extracted and expanded onto the harmonic series of
eq.~(\ref{eq:kinlin}), where the coefficients $(A_{n}, B_{n})$ are
determined by a least-squares fit with a basis $\{1, \cos\psi,
\sin\psi,...,\cos N\psi, \sin N\psi\}$.

The harmonic series can be presented in a more compact way,
\begin{equation}
  \label{eq:kinemetry}
  K(a,\psi) = A_{0}(a) +
            \sum_{n=1}^{N} k_{n}(a)\,\cos[n(\psi-\phi_{n}(a))].
\end{equation}

\noindent where the amplitude and phase coefficients
($k_{n}$, $\phi_{n}$) are easily calculated from the $A_{n}, B_{n}$
coefficients:
\begin{equation}
  \label{eq:coeff}
  k_{n}    =  \sqrt{A^2_{n} + B^2_{n}} \phantom{10} \textrm{and}  \phantom{10} 
  \phi_{n} =  \arctan \left( \frac{A_{n}}{B_{n}}\right).
\end{equation}

The $\sin$ and $\cos$ terms have different properties and, in
principle, describe different characteristics of the maps. The
properties of the corresponding coefficients have been studied and
described for the case of thin gaseous disks
\citep{1994ApJ...436..642F,1997MNRAS.292..349S,2004ApJ...605..183W}.
In a general case however, namely for the analysis of kinematic maps
of triaxial (stellar) systems, those properties do not apply (e.g. the
$b_1$ term represents circular rotation only in the thin disk
approximation). For these reasons, we find it more instructive in this
paper to combine the coefficients of the same order and characterise
the global properties of kinematic maps. In what follows, although the
implementation of the method determines individual coefficients (and
can be straight-forwardly used for the special cases of disks), they
will be presented in the spirit of eq.~(\ref{eq:kinemetry}).

%%%%%% Figure 1%%%%%%%%%%%%%%%%%%%%%%%%%%%%%%%%%%%%%%%%%%%%%%%%%%%%
\begin{figure*}
        \includegraphics[width=0.9\textwidth]{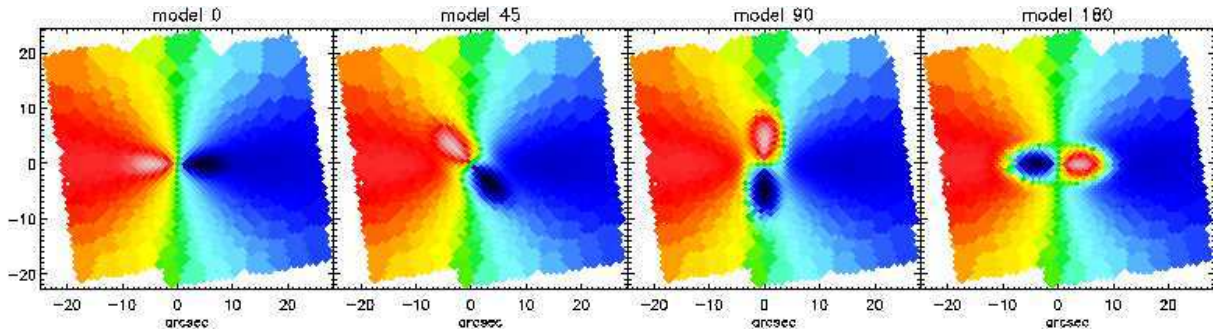}
  \caption{\label{f:models} 
    Four model maps representing different but typical velocity maps
    of triaxial early-type galaxies. Velocity maps were constructed of
    two components (see text for details). The position angle of the
    first (nuclear) component changes from left to right: 0, 45, 90
    and 180\degr. The size of bins increases towards the edge of the
    maps, as expected form observations, to keep constant
    signal-to-noise ratio. Note also the different opening angles of
    the two components on the maps. }
\end{figure*}
%%%%%%%%%%%%%%%%%%%%%%%%%%%%%%%%%%%%%%%%%%%%%%%%%%%%%%%%%%%%%%%%%%%%%%

\subsection{Internal error estimates}
\label{ss:error}

The uncertainties of the resulting parameters can be estimated from
the measurement errors of the input kinematic data. The ellipse
parameters determined by the Levenberg-Marquardt least-squares
minimisation code have formal one-sigma uncertainties computed from the
covariance matrix. If an input parameter is held fixed, or if it
touches a boundary, then the error is reported as zero. Similarly, the
harmonic terms obtained in the linear least-squares fit in the second
part of the procedure have their formal one-sigma errors estimated from
the diagonal elements of the corresponding covariance matrix, as
described in Section~15.6 of \citet{1992nrfa.book.....P}.

We performed Monte Carlo simulations to estimate the robustness of the
obtained internal errors. We used the {\tt SAURON}
\citep{2001MNRAS.326...23B} velocity map of NGC~2974 presented in
\citet{2004MNRAS.352..721E} and its measurement errors to construct
one hundred realisations of the velocity map.  The velocity in each
bin was taken from a Gaussian distribution of mean equal to the
observed velocity and standard deviation given by the corresponding
measurement error.  Typical errors on the velocities of the {\tt
  SAURON} observations are about $5 \kms$.  All realisations provide a
distribution of values from which we estimate one-sigma errors. The
Monte Carlo uncertainties are in a good agreement with the formal
errors. Note, however, that these errors are statistical errors only
and do not include systematic effects.  We investigate the resulting
accuracy of the kinemetric coefficients in Section~\ref{s:appli}.

%%%%%%%%%%%%%%%%%%%%%%%%%%%%%%%%%%%%%%%%%%%%%%%%%%%%%%%%%%%%%%%%%%
%
%SECTION4 SECTION4 SECTION4 SECTION4 SECTION4 SECTION4 SECTION4 
%
%%%%%%%%%%%%%%%%%%%%%%%%%%%%%%%%%%%%%%%%%%%%%%%%%%%%%%%%%%%%%%%%%%

\section{Kinematic parameters and their meaning}
\label{s:kinpar}

The crucial aspect of any analysis method is the interpretation of its
results. In this section, we describe in more detail the way the
method works and demonstrate the meaning of the resulting kinemetric
parameters.

\subsection{Model velocity maps}
\label{ss:mmaps}

We constructed a set of model velocity maps combining two components
using the circular velocity from the \citet{1990ApJ...356..359H}
potential, weighted with two different surface brightness
distributions. The Hernquist potential was chosen because it
approximates reasonably well the density of real early-type galaxies.
Its circular velocity is given by $V_c = \sqrt{GMr}/(r+r_0)$, where
$G$ is the gravitational constant, $M$ is total mass of the system,
and $r_0$ is scale length of the potential.  The $\sqrt{GM}$ factors
for the two components were 850 and 1500 (in units of $\kms$
arcsec$^{1/2}$), respectively, while $r_0$ scale lengths were set to
5\arcsec and 15\arcsec, defining a central and large scale component,
respectively.

The light distribution of the first component was described by an
exponential law typical of disks, and that of the second component by
an $r^{1/4}$ law typical of spheroids. They were both normalised to a
maximum value of unity. The light distribution scale lengths, $r_e$,
of the two components were set to 7\arcsec and 15\arcsec,
respectively.  The components were inclined (60\degr and 40\degr,
respectively) and combined with 4:1 relative weights, respectively,
rotating the central component for different position angles $\Gamma =
(0\degr, 45\degr, 90\degr, 180\degr)$ in order to construct different
features on velocity maps possible to occur in triaxial early-type
galaxies.

The final result are four model velocity maps, which satisfy the
symmetry relations of eq.~(\ref{eq:paspolar}), presented in
Fig.~\ref{f:models}. Each model has a distinct nuclear component with
a somewhat higher maximum velocity and a smaller opening angle than
the large scale, spheroidal component. We define the opening angle
loosely as a quantity that describes the tightness of the iso-velocity
contours. It is discussed in more detail below.  The main difference
between the models is in the orientation of the nuclear component.
All model maps were constructed to simulate realistic observations
with the {\tt SAURON} integral-field spectrograph, where the size of
the spatial bins increases from the centre outwards to maintain
constant signal-to-noise ratio.

%%%%%% Figure 2%%%%%%%%%%%%%%%%%%%%%%%%%%%%%%%%%%%%%%%%%%%%%%%%%%%%
\begin{figure*}
        \includegraphics[width=\textwidth]{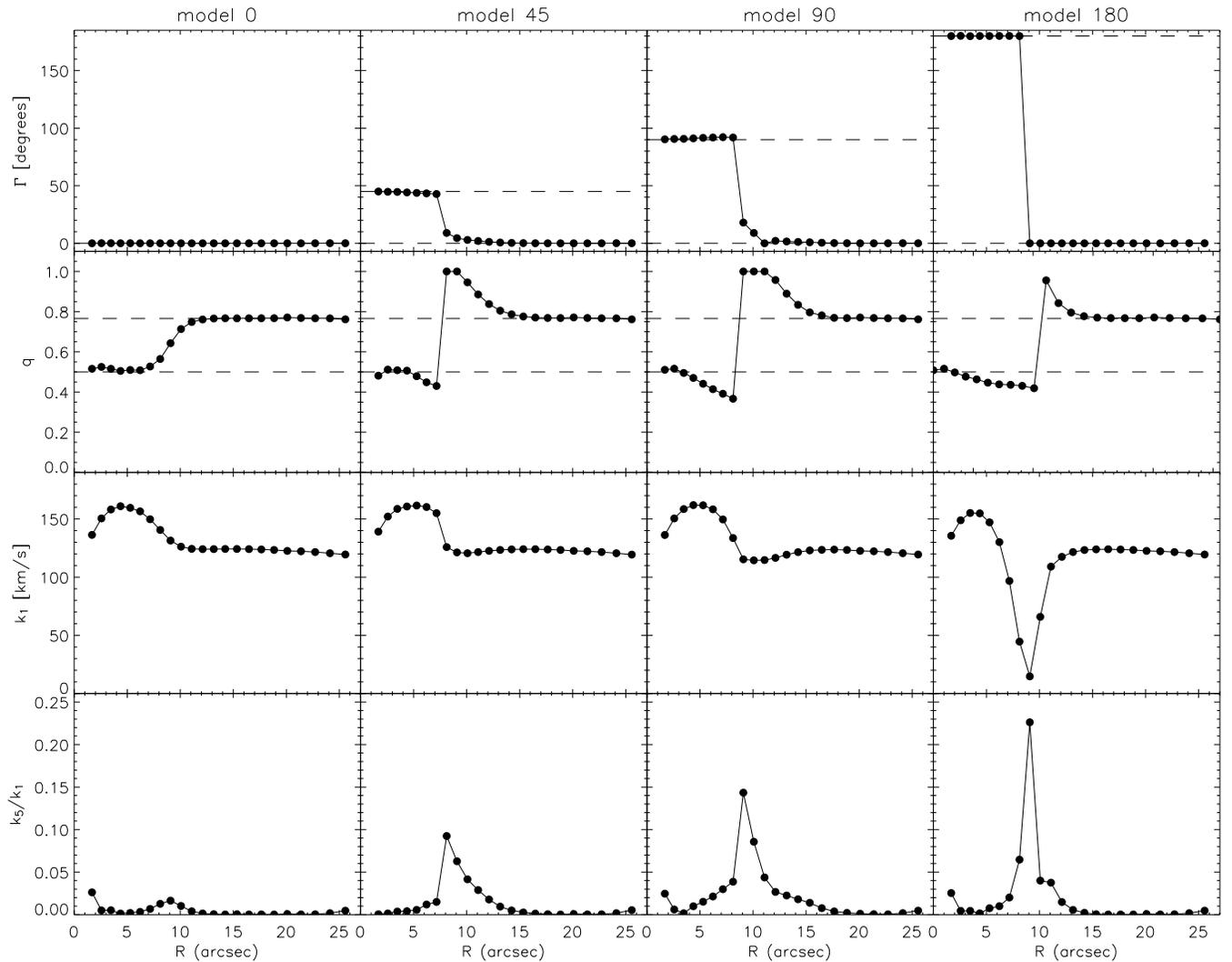}
  \caption{\label{f:coeffs} 
    Kinemetric coefficients of the model maps. From left to right the
    plots show the results for {\it model 0}, {\it model 45}, {\it
      model 90} and {\it model 180}. From top to bottom: kinematic
    position angle $\Gamma$, flattening $q$, coefficients of harmonic
    expansion $k_1$ and $k_5/k_1$. Note that the sudden rise of
    $k_5/k_1$ in {\it model 180} at 9\arcsec is due to the nearly zero
    value of the rotation velocity $k_1$ at that radius.}
\end{figure*}
%%%%%%%%%%%%%%%%%%%%%%%%%%%%%%%%%%%%%%%%%%%%%%%%%%%%%%%%%%%%%%%%%%%%%%

\subsection{Kinemetric analysis of model maps}
\label{ss:kinanal}

The parity of an odd kinematic moment guarantees that the even terms
in the expansion vanish in the absence of measurement errors and
assuming that the centre is given correctly. These conditions are
satisfied for the model velocity maps and, while applying the
algorithm of section~\ref{ss:alg}, we simplified
eqs.~(\ref{eq:kinNew}) and ~(\ref{eq:cchi2}) by not minimising the
contribution of $A_2$ and $B_2$.  However, during the harmonic
analysis we expanded the extracted velocity profiles including two
additional odd harmonic terms, $A_5$ and $B_5$, which describe the
deviations from circular rotation (see eq.~\ref{eq:tilt}).

A special case is that of the zeroth-order term given by the $A_0$
coefficient. It is not shown in the subsequent analysis because the
absolute level of the maps was set to zero by construction. For
velocity maps, $A_0$ gives the systemic velocity of each ring.  In the
case of even kinematic moments, the meaning of $A_0$ is different: it
is the dominant term describing the amplitude of the moment (e.g.
intensity for surface brightness, amplitude for velocity dispersion of
velocity dispersion maps, see section~\ref{ss:jeans}).

Figure~\ref{f:coeffs} shows the kinemetric parameters which describe
the model velocity maps. All parameters are plotted versus the
semi-major axis length of the ellipses. The first and second row show
the orientation ($\Gamma$) and flattening ($q$) of the ellipse along
which the velocity profile was extracted.  These parameters uniquely
specify the best fitting ellipse, along which the contribution of
$A_1$, $A_3$ and $B_3$ terms is minimal. The following rows show plots
of the harmonic coefficients given in the compact form of
eq.~(\ref{eq:coeff}): $k_1$ and $k_5/k_1$.

\begin{figure}
        \includegraphics[width=\columnwidth]{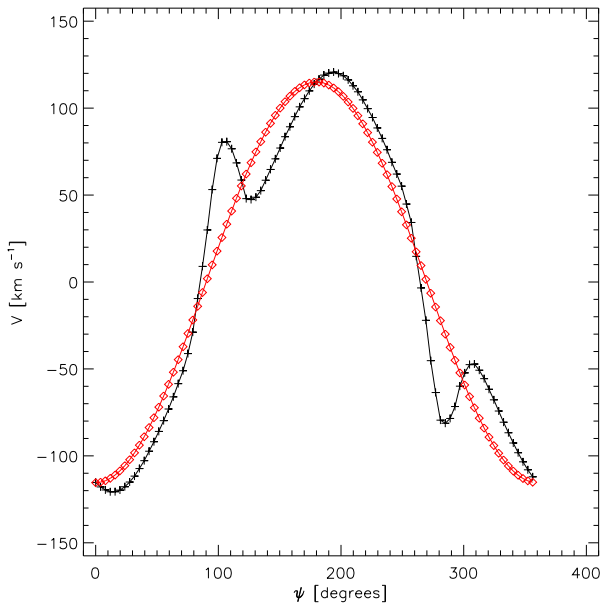}
  \caption{\label{f:prof} 
    Velocity profile taken from the {\it model 90} map along a circle
    ($q=1$) at r=9\arcsec, in the transition region between the two
    components. Overplotted with diamonds is the circular velocity
    profile at the same position. The observed profile exhibits a
    complex structure which can not be represented by requiring that
    the third-order terms in the harmonic expansion be zero. The local
    extremes at $\psi \approx 100\degr$ and $280\degr$ originate in
    the nuclear component, while the global maximum and minimum are
    given by the large scale component.  }
\end{figure}

The meaning of these parameters can be associated with the visible
structures on the maps as follows:

\begin{enumerate}
  
\item $\Gamma$ is the kinematic position angle. Like the photometric
  position angle, it describes the orientation of the velocity map,
  and it is related to the projection of the angular momentum vector
  (see Appendix~\ref{s:kinpa}).  $\Gamma$ traces the position of the
  maximum velocity on the map.
  
\item $q$ is the axial ratio or flattening of the ellipse and it
  describes the flattening of the map. It can be interpreted as a
  measure of the opening angle of the iso-velocity contours. Maps that
  have large opening angles appear {\it round} (large values of $q$),
  while maps with small opening angle appear {\it flat} (small values
  of $q$). In the special case of a disk, where the motion is confined
  to a plane (spiral galaxies, gaseous disks, etc), the flattening is
  directly related to the inclination of the disk, $q = \cos(i)$, where
  the projected ellipses correspond to intrinsically circular orbits.
  
\item $k_1$ is the dominant coefficient on maps of odd kinematic
  moments. For velocity maps, it describes the amplitude of bulk
  motions (rotation curve).  It contains contributions from both
  $\cos \psi$ and $\sin \psi$ terms, and in the special case of a
  thin disk geometry, the $A_1$ and $B_1$ terms should be considered
  separately. In the case of perfect circular motions ($A_1$ and other
  higher terms are zero), $k_1$ describes the full motion on the map.
  In other words, if higher order coefficients are zero (or
  negligible), then the velocity map is identical to the velocity
  map of a disk viewed at inclination $i$.
  
\item $k_3$ is the combination of harmonic coefficients that are
  minimised in the fitting process. This coefficient is sensitive to
  the, centre, flattening and position angle of the ellipse and in
  special cases it will be larger then zero indicating a velocity
  profile with a complex structure (multiple extremes) originating
  from the overlap of multiple kinematic components with different
  orientations (see Fig.~\ref{f:prof}). This coefficient, however,
  does not bring any additional information to $q$ and $\Gamma$ and we
  do not show it. On the other hand, if $q$ and $\Gamma$ are fixed and
  not fitted, as is often the case in studies of disk velocity maps,
  then this term brings the first correction to the simple rotational
  motion.
  
\item $k_5$ is the first higher-order term which does not define the
  parameters of the best fitting ellipse. It represents the
  higher-order deviations from simple rotation and points to complex
  structures on the maps. This term is thus sensitive to the existence
  of separate kinematic components. It is a kinemetric analogous of the
  photometric term that describes the deviation of the isophote shape
  from an ellipse (diskiness and boxiness)

\end{enumerate}

%%%%%% Figure 3%%%%%%%%%%%%%%%%%%%%%%%%%%%%%%%%%%%%%%%%%%%%%%%%%%%%
%%%%%%%%%%%%%%%%%%%%%%%%%%%%%%%%%%%%%%%%%%%%%%%%%%%%%%%%%%%%%%%%%%%%%%

Comparison of the structures on the model maps (Fig.~\ref{f:models})
and the radial dependence of the kinemetric coefficients
(Fig.~\ref{f:coeffs}) illustrates the descriptive power of the method.
Our model maps were built of two components which had different
flattening and kinematic position angle. It is the role of the
kinemetric coefficients to recognise these characteristics and
describe them (Fig.~\ref{f:coeffs}).

The components of {\it model 0} were aligned, but the nuclear
component iso-velocities had a small opening angle, while those of the
large-scale component had a large opening angle. This is reflected in
the different flattenings $q$. In the inner 10\arcsec , the disky
component dominates and the map is rather {\it flat}, while beyond
this region the spheroidal component dominates and the map looks more
{\it round}. In the regions where only one component dominates, the
recovered axial ratio corresponds correctly to the inclination of that
component. While the $k_1$ terms show the amplitude of the rotation,
the $k_3$ and $k_5$ coefficients are close to zero.  This is expected
because the models were built of two maps of essentially circular
velocity. A small $k_5$ terms exists in the {\it transition region}
between the components. This suggests that the combination of aligned
but separate velocity components will give rise to non-zero $k_5$.

A confirmation of this effect is given by {\it model 180}, where the
two components are aligned but counter-rotating. In this configuration
the components are clearly visible on the $q$ and $k_1$ panels and
$k_5$ rises strongly in the region coinciding with the change of
rotational sense (change in the kinematic position angle by 180\degr).

Differences in kinematic position angle between the components, as in
the maps of {\it model 45} and {\it model 90}, generate more power in
higher order coefficients. The position angle is recovered correctly
in the regions dominated by one component, while in the region between
the components it traces the position of (combined) maximum velocity.
It is the same with the flattening, which follows the change of the
map topology, reaching extreme round values in the transition region
where both components equally dominate.  The reason for this is the
complex shape of the velocity profile, for which it becomes impossible
to make the $A_3$ and $B_3$ terms along any ellipse equal to zero
(Fig.~\ref{f:prof}).  In such a case, the fitting routine chooses the
equal ratio of axes as this is the most general case, confirming the
robustness of the approach. The two stage fit, direct grid fitting
followed by non-linear minimisation, is necessary to provide the
required robustness in such regions.

The extent of the transition region depends on the relative position
angle, larger angles resulting in a larger transition region. This is
reflected in the behaviour of the higher-order terms, which become
more important and influence a larger area with increasing
misalignment of the components. The increasing misalignment is
effectively separating the components, creating a complex,
multi-peaked velocity profile such as in Fig.~\ref{f:prof}. Note that
strong contribution of higher-order terms in this examples is not
necessarily representative of real galaxies, as the features on the
velocity maps were not modelled to represent specific observed cases.

\begin{figure*}
        \includegraphics[width=0.95\textwidth]{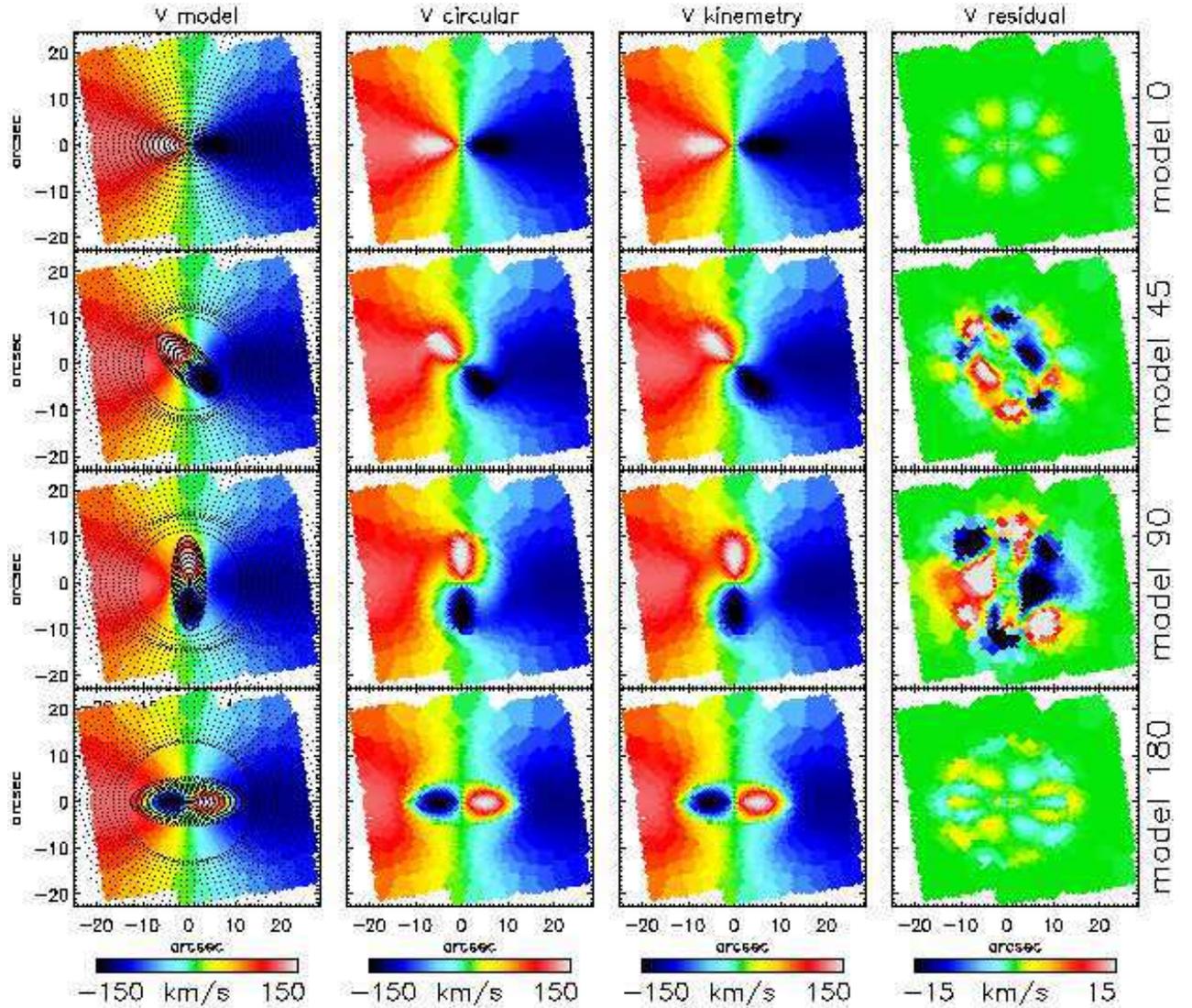}
  \caption{\label{f:maps} 
    Model velocity maps and the results of the kinemetric analysis.
    From top to bottom: maps corresponding to {\it model 0}, {\it
      model 45}, {\it model 90} and {\it model 180} (as noted at the
    end of each row).  From left to right: model maps, maps of
    circular velocity, reconstructed model maps using 6 harmonic
    terms, residual between circular and reconstructed velocity maps.
    Overplotted on the model maps are the best fitting ellipses along
    which the velocity profiles were extracted and analysed. Note the
    change in flattening and position angle of the ellipses.}
\end{figure*}

Figure~\ref{f:maps} shows the two-dimensional representation of the
analysis. For each model, we reconstructed a map using only the $b_1$
term which is not minimised, and we named those new maps 'circular
velocity'. We also reconstructed a map using all 6 terms of the
harmonic expansion, which we named 'kinemetric velocity'. The
difference between the two velocity maps shows the contribution of the
higher order terms ($k_3$ and $k_5$). These residual maps of {\it model
  0} and {\it model 180} show distinct five-fold symmetries coming
from the high $k_5$ term. Residuals of other models show a combination
of the $k_3$ and $k_5$ terms in the transition region where velocity
profiles are similar to the one in Fig.~\ref{f:prof}

The two examples with five-fold symmetries are interesting when
compared with the corresponding model maps. Both residual maps have
positive amplitudes on the negative side of the major axis. This is,
however, not the case for the model maps. The {\it model 0} map has
two components which both rotate in the same direction, while for {\it
  model 180} the inner component counter-rotates with respect to the
outer component. Both models have a strong $k_5$ term, and hence
five-fold residual structure in the transition region. Interestingly,
the amplitude of the residuals of {\it model 180} at e.g. $r=7\arcsec$
along the major axis is negative while the mean rotation at the same
radius is positive, belonging to the region dominated by the
counter-rotating component. It is thus possible to infer that the
residuals actually trace the contribution of the main body rotation in
this region, although this is not visible on the velocity map. A
similar situation is visible in {\it model 0}, but the two components
co-rotate and the residuals show the same sense of rotation.

We conclude this exercise stating that $k_5$ is sensitive to the
existence of different components, where the increase of $k_5$ in the
presented models is correlated to larger misalignments between the
components.

%%%%%% Figure 4%%%%%%%%%%%%%%%%%%%%%%%%%%%%%%%%%%%%%%%%%%%%%%%%%%%%
%%%%%%%%%%%%%%%%%%%%%%%%%%%%%%%%%%%%%%%%%%%%%%%%%%%%%%%%%%%%%%%%%%%%%%

\subsection{Special cases}
\label{ss:special}

There are two possible cases of velocity maps for which the
determination of kinemetric coefficients will be degenerate. The first
case arises from the effect of solid body rotation. As shown in
\citet{1997MNRAS.292..349S}, while it is possible to measure the
position angle $\Gamma(a)$, it is impossible to uniquely determine the
flattening $q(a)$, the centre and the systemic velocity. This
degeneracy is present because all iso-velocity contours are parallel,
and the opening angle is $180\degr$.

Another extreme case occurs when rotation is very slow. Such velocity
maps are often found in giant ellipticals which are thought to be
triaxial bodies. Here the velocity profiles are roughly constant and
do not show the parity expected from an odd moment of the LOSVD.
There is no gradient in velocity and as a consequence the flattening
and position angle cannot be determined.

In these two cases the new method cannot be applied blindly because it
is not possible to interpret harmonic terms in the way we showed up to
now. There are two possible ways to proceed. One way is to make an
assumption about the flattening and the kinematic position angle (e.g.
use photometric flattening), and continue with the harmonic analysis
of the velocity profiles extracted along those ellipses.

A more general way, for both cases, is to extract the velocity
profiles along concentric circles and perform the harmonic analysis on
those. In this way one does not bring any assumption in the analysis,
but it is still possible to determine the rotation velocity and
kinematic position angle (if present) and detect trends in
higher-order harmonic terms. A detailed discussion of extraction along
circles and the meaning of the coefficients is given in
Appendix~\ref{s:circ}. An automatic transition towards circles can be
implemented by adding a penalty term to the $\chi^2$, to bias the
solution towards round shapes, when the gradient is small (or
iso-velocities are parallel). The goodness of fit is then determined
by $\chi^2_p = \chi^2 + \alpha P(q)$. We also note that an expansion
on circles might be good for the analysis of higher-order moments of
the LOSVD which normally have rather poor signal-to-noise ratio.

%%% Figure 5%%%%%%%%%%%%%%%%%%%%%%%%%%%%%%%%%%%%%%%%%%%%%%%%%%%%%%%%%%%%%%%

\begin{figure*}
  {\hbox {\epsfxsize=0.33\textwidth
      \epsfbox{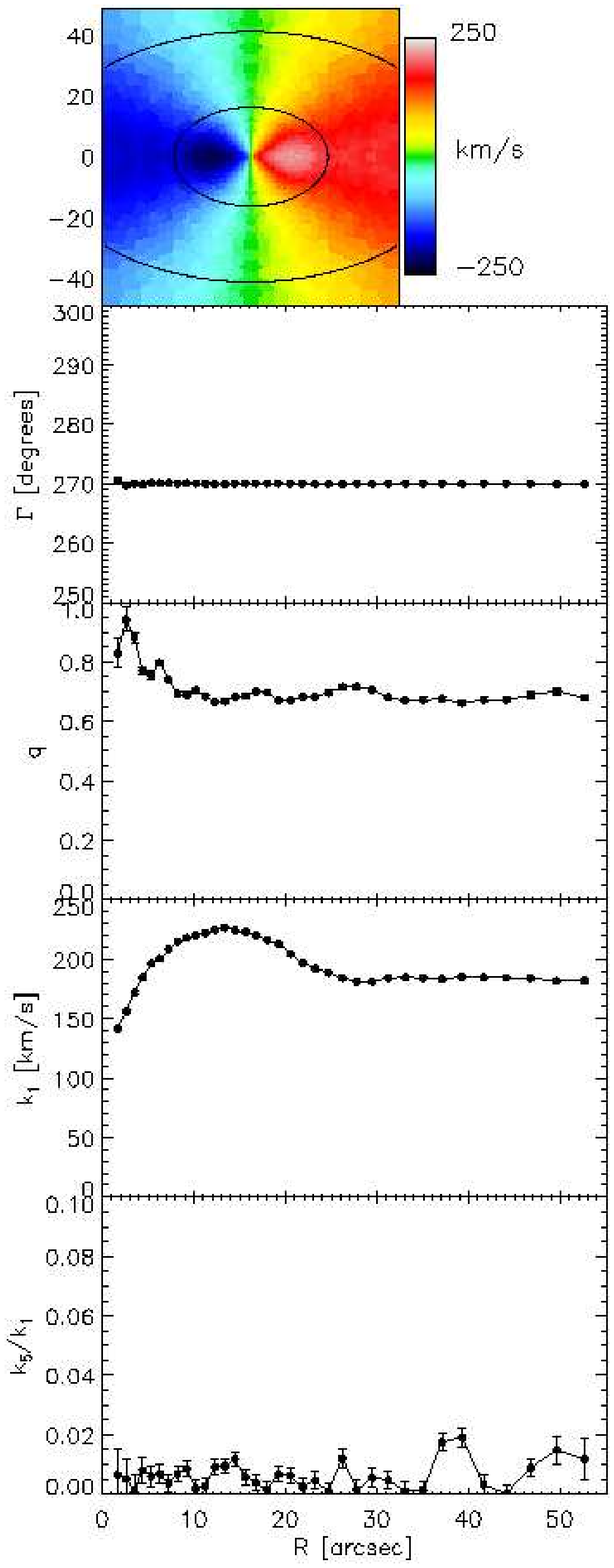}
       \epsfxsize=0.34\textwidth
      \epsfbox{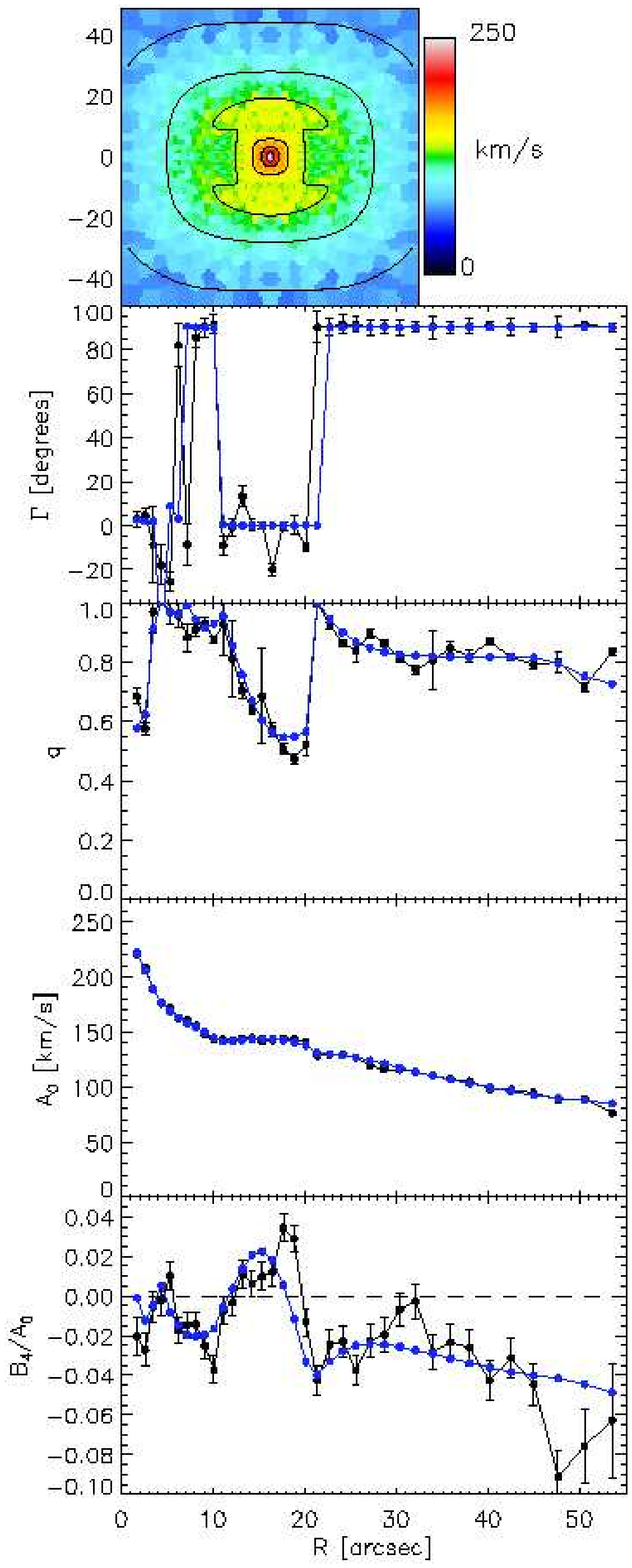}
       \epsfxsize=0.33\textwidth
      \epsfbox{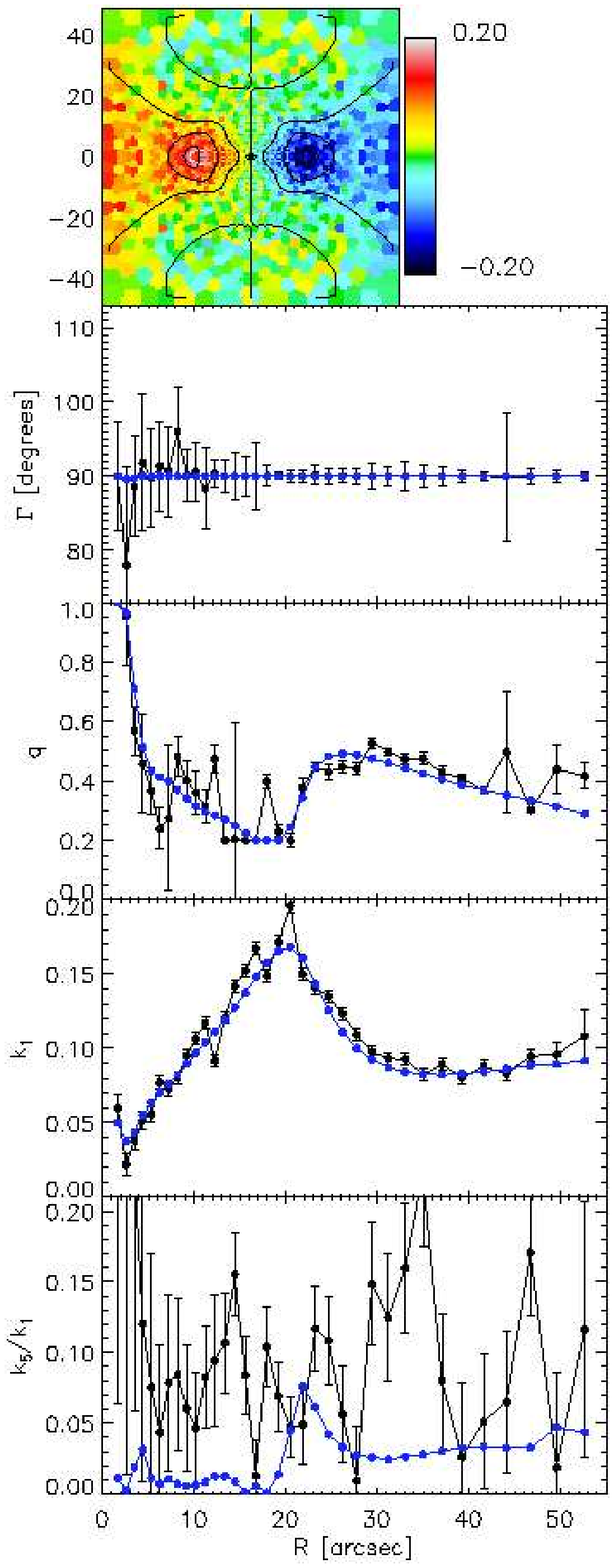}
      }}
\caption{\label{f:2Imodel}
  Maps of mean velocity (right), velocity dispersion (middle)
  $\sigma$, and Gauss-Hermite moment $h_3$ (left) for the model galaxy
  with added intrinsic scatter corresponding to a realistic
  measurement with {\tt SAURON}. The overplotted ellipses on the
  velocity map are the surface brightness contours. The plots in each
  right and left column represent the kinemetric parameters ($\Gamma$,
  $q$, $k_1$, $k_5/k_1$) describing the maps.  The plots in the middle
  column are kinemetric analogs of the standard photometric
  parameters: position angle, flattening, intensity ($A_0$ term) and
  the shape parameter ($B_4$).  Horizontal line in the last plot of
  the middle column marks the division between boxiness and diskiness
  of iso-$\sigma$ contours.  Blue lines in the middle and right column
  plots correspond to the noiseless maps of $\sigma$ and $h_3$, whose
  contours are overplotted on the $\sigma$ and $h_3$ map.}
\end{figure*}
%%%%%%%%%%%%%%%%%%%%%%%%%%%%%%%%%%%%%%%%%%%%%%%%%%%%%%%%%%%%%%%%%%%%%%%%%%

%%%%%%%%%%%%%%%%%%%%%%%%%%%%%%%%%%%%%%%%%%%%%%%%%%%%%%%%%%%%%%%%%%
%
%SECTION5 SECTION5 SECTION5 SECTION5 SECTION5 SECTION5 SECTION5 
%
%%%%%%%%%%%%%%%%%%%%%%%%%%%%%%%%%%%%%%%%%%%%%%%%%%%%%%%%%%%%%%%%%%

\section{Application}
\label{s:appli}

In this section we apply our kinemetry formalism to three different
examples.  Our wish is to justify the choice of expansion on ellipses
along which the profiles satisfy the simple cosine law. We first apply
the method to the odd and even kinematic moments of an axisymmetric
model of a galaxy. Then we apply the method to a few observed
galaxies. Finally, we offer a characterisation of structures on the
velocity maps which can be recognised through the behaviour of the
kinemetric parameters.

\subsection{Axisymmetric two-integral model}
\label{ss:jeans}

The working assumption of the method introduced in
Section~\ref{s:back} is that there exists an ellipse along which the
profile of a kinematic moment can be described by a simple cosine law
(eq.~\ref{eq:tilt}). This profile is easily associated with the
circular motions in a thin disk \citep[e.g.][]{1997MNRAS.292..349S},
which show regular velocity maps.  More general stellar systems with
axial symmetry, however, also show very regular velocity maps. Here we
demonstrate that our working assumption of eq.~(\ref{eq:tilt}) is
satisfied in realistic models for axisymmetric galaxies.

We constructed an axisymmetric model galaxy with a distribution
function which depends only on the energy and one component of the
angular momentum. The model was constructed using the
\citet{1993MNRAS.262..401H} contour integration as in method
\citet{1999MNRAS.303..495E} to resemble the \SAURON observations of
NGC~2974 \citep{2004MNRAS.352..721E, 2005MNRAS.357.1113K}. The
distribution function of the model yields the full LOSVD from which
the observable kinematics can be extracted.  This was done on a large
$100\arcsec \times 100\arcsec$ field by fitting a Gauss-Hermite series
($V$, $\sigma$, $h_3$ and $h_4$) to spatially binned data resembling
typical \SAURON observations. In order to mimic real observations, we
also assigned typical \SAURON observational errors to each model
observable (the relevant errors here are on velocity, velocity
dispersion and $h_3$, respectively $5 \kms$, $7 \kms$ and 0.03). These
errors were used to add intrinsic scatter to the noiseless model data
by means of a Monte Carlo realisation. For further details of the
model, we refer the reader to Section~5.1 of
\citet{2005MNRAS.357.1113K}.  Model maps of velocity, velocity
dispersion and $h_3$ are presented in Fig.~\ref{f:2Imodel}.

We performed a kinemetric analysis of two odd and one even moment of
the model galaxy. The left column of Fig.~\ref{f:2Imodel} presents the
results for the velocity map. The flattening is constant over most of
the map with a small but continuous change in the centre.  The
kinematic position angle is very well determined and remains constant.
The first harmonic term, $k_1$, describes the rotational motion in the
model galaxy and peaks around 15\arcsec. The higher order term $k_5$,
mostly consistent with zero, shows some deviations on a 0.5-1\% level
in the centre and slightly more at larger radii.

The middle column of Fig.~\ref{f:2Imodel} presents the analysis of the
first even moment of the LOSVD: the velocity dispersion map. As
mentioned earlier, because even moments have the same parity as the
surface brightness (zeroth moment), the kinemetric analysis of an even
moment, such as velocity dispersion, is analog to the photometric
analysis of the surface brightness. We demonstrate this by fitting the
velocity dispersion map using our code, but with the requirement that
$\mu_{e}(a, \psi)= A_{0}(a)$ along a trial ellipse. This yields the
usual photometeric parameters: position angle, ellipticity, intensity
and the shape parameters describing the deviations from the ellipse.
In our case of an even kinematic moment we recognise them as the:
kinematic position angle $\Gamma$, flattening ($q=1-\epsilon$), $A_0$
and $B_4$ harmonic terms, while other harmonic terms in
eq.~(\ref{eq:kinlin}) are zero. The zeroth harmonic term, $A_0$, is
the dominant term in the expansion and it represents velocity
dispersion profile with radius. The fourth harmonic term, $B_4$, can
be used to describe the deviations of the iso-$\sigma$ contours from
an ellipse, where, as in photometry, negative values are indication of
boxiness, while positive values of diskiness. The given velocity
dispersion map is in general boxy, but between 10\arcsec and 20\arcsec
it becomes disky. The same region is marked by a change in flattening
(elipticity) and orientation (position angle).

Observed velocity dispersion maps are usually noisier than the
velocity maps. We analysed a noiseless model velocity dispersion map
(without intrinsic scatter) to determine with what confidence one
could analyse an observed velocity dispersion map. The results are
also plotted in the middle column of Fig.~\ref{f:2Imodel}, and they
show that the main features of kinemetric coefficients are well
recovered, suggesting that the method could be used to analyse
velocity dispersion maps obtained from data with high signal-to-noise
ratio.

The right column of Fig.~\ref{f:2Imodel} presents the results of a
similar exercise, but this time on the observed third moment of the
LOSVD (given by Gauss-Hermite parameter $h_3$). The noise on this map
considerably distorts the structure of the map. The kinematic position
angle is again very well determined, being, as expected, offset
180\degr\, from the kinematic angle of the velocity map, and the $k_1$
term traces well the amplitude of the $h_3$ with radius.  The
flattening changes with a jump at $\approx25\arcsec$. The higher-order
coefficients are however substantially influenced by the noise. The
$k_5$ term is often more than 10 \% of $k_1$.  We repeated the
analysis on the noiseless (no intrinsic scatter) map of $h_3$ moment.
The new values of $q$, PA and $k_1$, also shown in
Fig.~\ref{f:2Imodel}, are in a good agreement with the values
extracted from the map with the intrinsic scatter.  This is, however,
not the case for the last term, $k_5$, where the noiseless model term
is small (consistent with zero) everywhere except in the region of the
change of the flattening $q$.  This feature is completely absent from
the noise dominated $k_5$ term of the intrinsic scatter map. We
conclude from this exercise that, although the method can be applied
straight-forwardly to the higher moments of the LOSVD, the
signal-to-noise ratio in current state of the art measurements is
generally not high enough to reliably recover kinemetric parameters of
the kinematic moments beyond velocity dispersion.

The negligible higher-order harmonic terms for both odd moments
confirm that the assumption of a simple cosine law for the kinematic
moments is justified in the case of an axisymmetric galaxy. In other
words, the method assumes that to zeroth order galaxies are
axisymmetric rotators. The non-zero higher-order harmonic terms
suggest the presence of multiple components ($k_5 \ne 0$). A change in
$q$ and $\Gamma$, however, hints to departures from
axisymmetry\footnote{Strictly speaking, in axisymmetric galaxies
  $\Gamma$ has to be constant and equal to the photometric position
  angle.}.

\subsection{{\tt SAURON} velocity maps}
\label{ss:ngc4473}

%%%% Figure 6%%%%%%%%%%%%%%%%%%%%%%%%%%%%%%%%%%%%%%%%%%%%%%%%%%%%%%%%%%%%%%
\begin{figure*} \centering
{\hbox { 
 \epsfxsize=0.24\textwidth \epsfbox{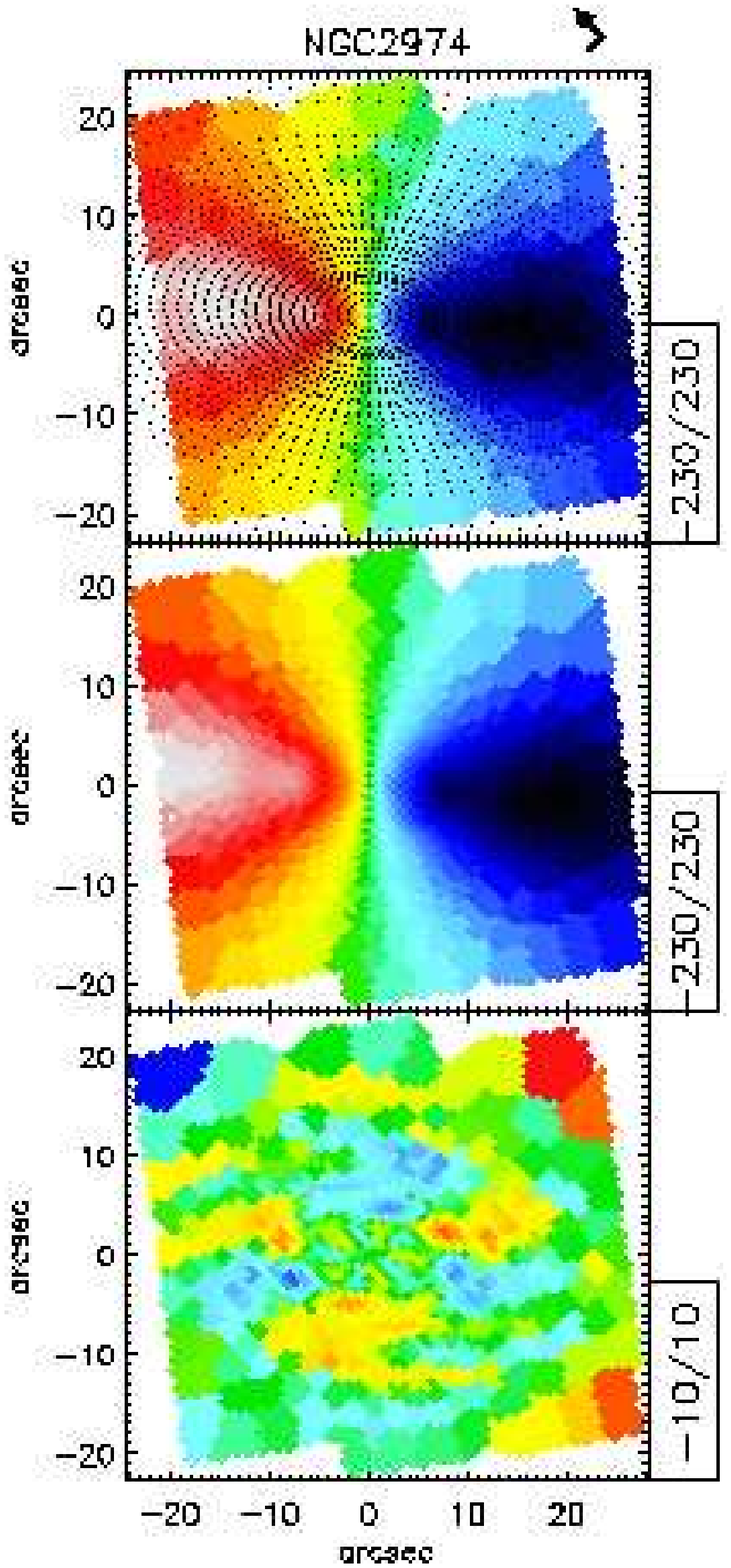} 
 \epsfxsize=0.23\textwidth \epsfbox{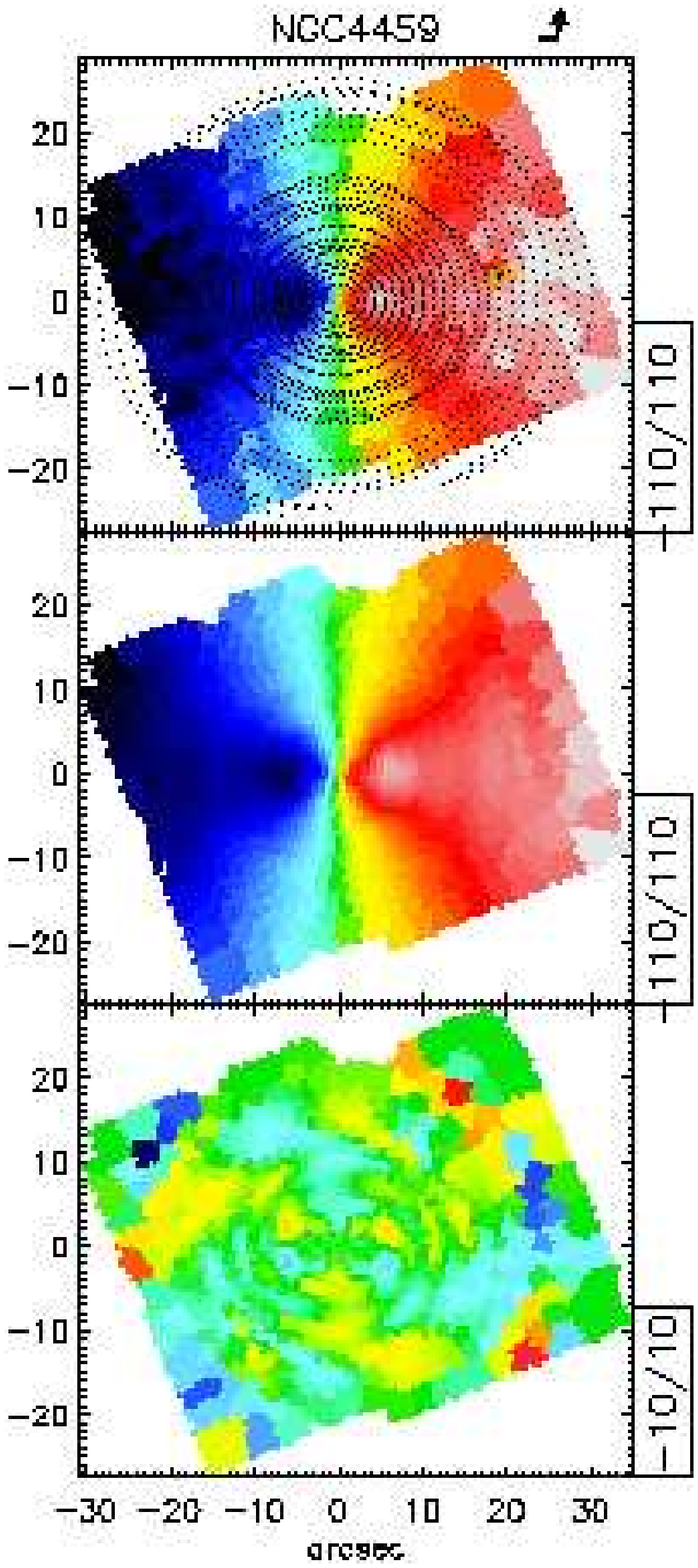} 
 \epsfxsize=0.252\textwidth \epsfbox{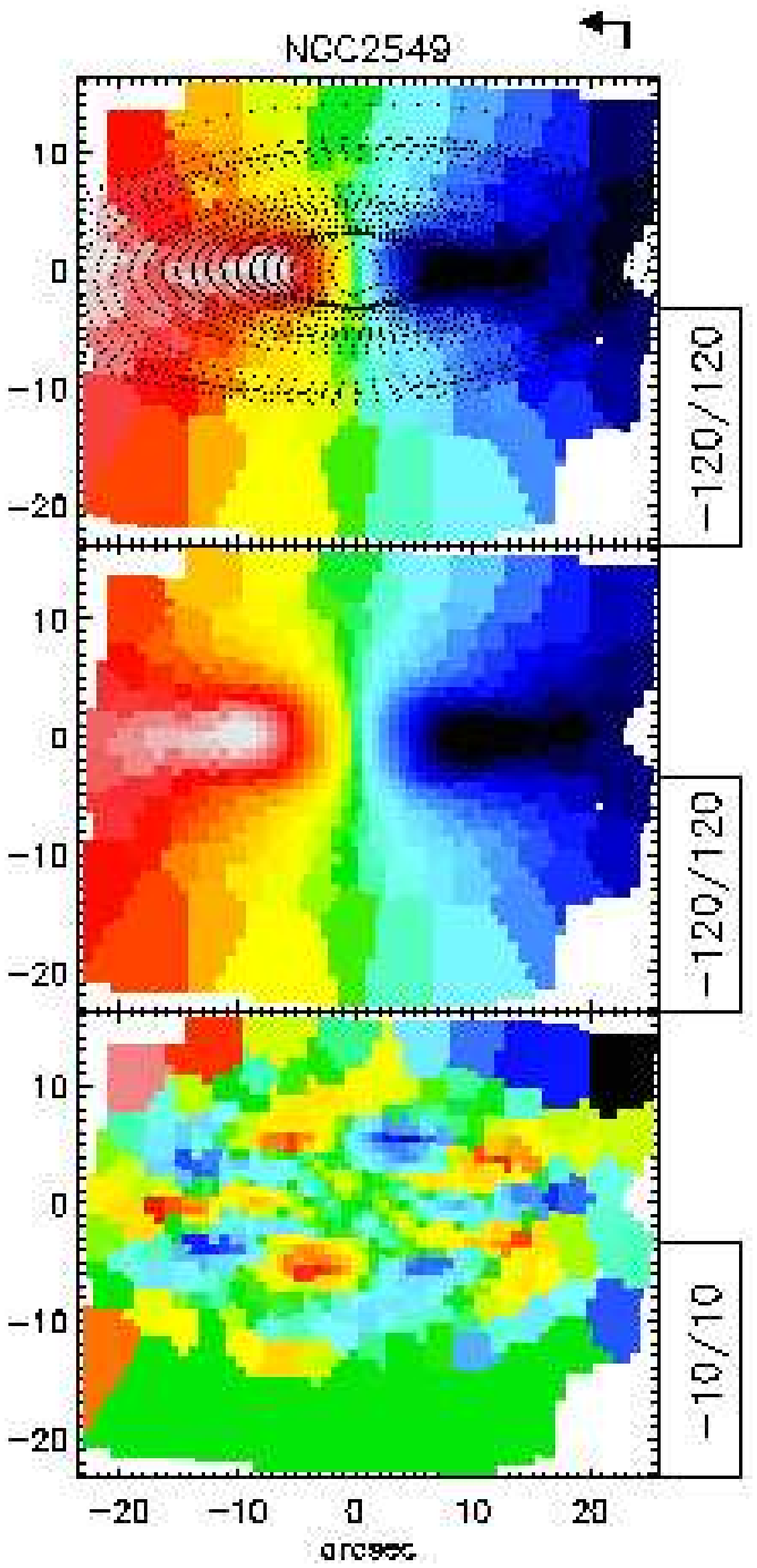}
 \epsfxsize=0.288\textwidth \epsfbox{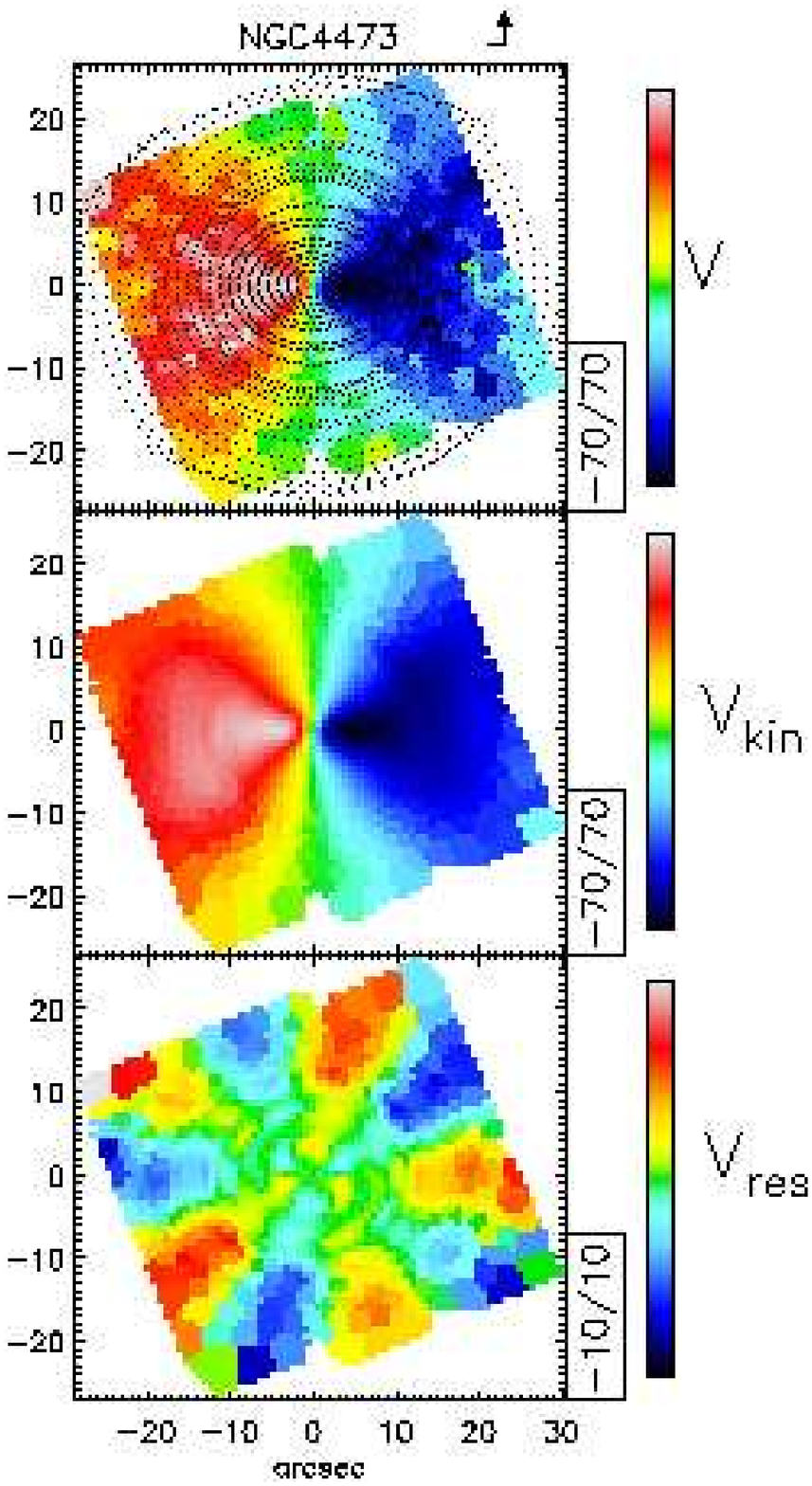} }}
\caption{\label{f:residuals}
  Velocity maps, reconstructed kinemetric velocity maps and the
  corresponding residual map (constructed as in Fig.~\ref{f:maps}) of
  NGC~2974, NGC~4459, NGC~2549 and NGC~4473 (from left to right). The
  best fitting ellipses are overplotted on the velocity maps. Maps
  were oriented such that the major axis is horizontal, and the
  north-east orientation of the maps is shown with the arrows.
  Residuals of NGC~2974 and NGC~4459 are smaller then the five-fold
  residuals of NGC~2549 and NGC~4473. Along the major axis, the
  residuals of NGC~2549 have the same sense of rotation as the
  velocity map, while residuals of NGC~4473 show the opposite sense of
  rotation.  }
\end{figure*}
%%%%%%%%%%%%%%%%%%%%%%%%%%%%%%%%%%%%%%%%%%%%%%%%%%%%%%%%%%%%%%%%%%%%%

We now apply the method to a few galaxies observed in the {\tt SAURON}
survey \citep{2002MNRAS.329..513D}. We selected the velocity maps of
NGC~2549, NGC~2974, NGC~4459 and NGC~4473 because they show diverse
velocity structures. The maps were presented in
\citet{2004MNRAS.352..721E} and are reproduced in the first row of
Fig.~\ref{f:residuals}. Inspection of the two-dimensional maps shows
that the iso-velocity contours of NGC~2549 behave differently in the
central and outer regions, suggesting the existence of two velocity
components. The velocity map of NGC~2974 is remarkably regular, with a
constant opening angle of the iso-velocities. NGC~4459 has a round
velocity map with a large opening angle, and the velocity map of
NGC~4473 shows an unusual decrease of the velocity along the major
axis.

Our purpose here is not to undertake a detailed study of the
properties of these galaxies, but to illustrate quantitatively the
application of the new method. Kinemetric profiles for the velocity
maps of the four galaxies are shown in Fig.~\ref{f:mosaic}. We also
show the photometric position angle and flattening (where
$q_{phot}=1-\epsilon_{phot}$) for a tentative comparison between the
surface brightness (as an even moment of the LOSVD) and velocity (an
odd moment of the LOSVD).  A detailed study of the photometric
properties of the {\tt SAURON} galaxies will be presented in a paper in
preparation by the {\tt SAURON} team.
%
%is presented in Damen et al. (in
%prep.).  
We summarise the main results as follows:

%%% Figure 7%%%%%%%%%%%%%%%%%%%%%%%%%%%%%%%%%%%%%%%%%%%%%%%%%%%%%%%%%%%%%%
\begin{figure*} \centering
{\hbox { 
 \epsfxsize=0.256\textwidth \epsfbox{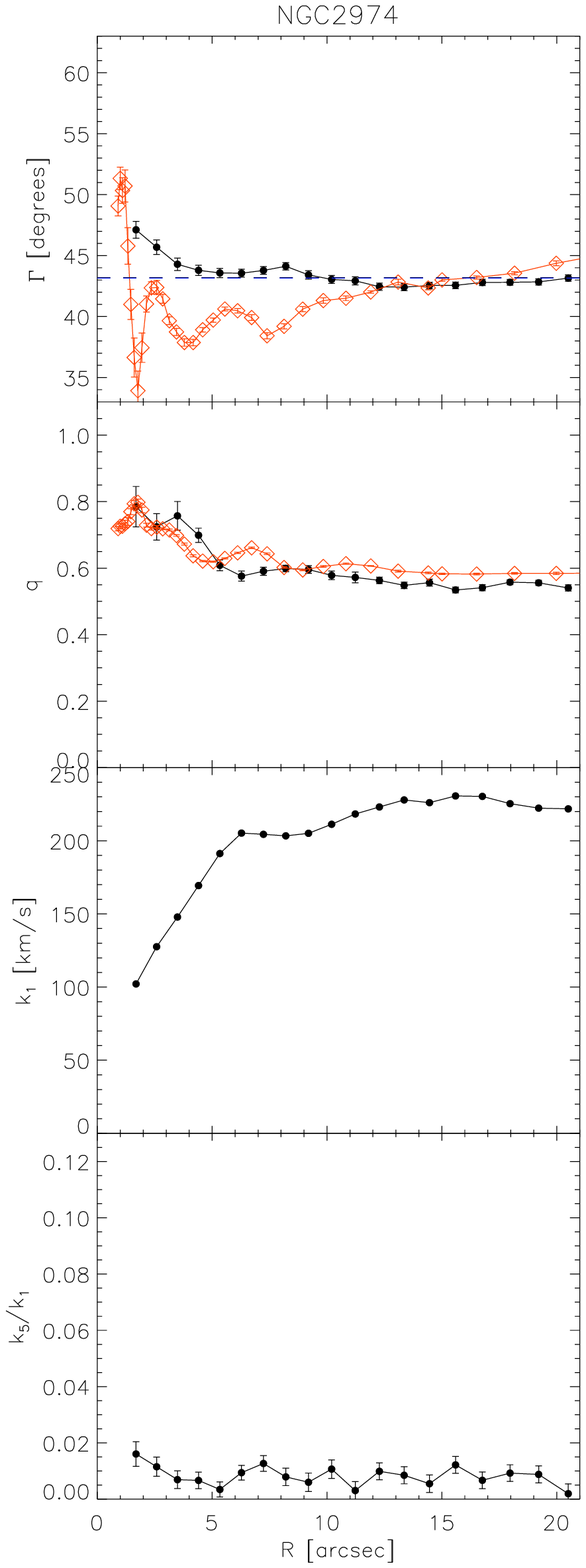} 
 \epsfxsize=0.24\textwidth \epsfbox{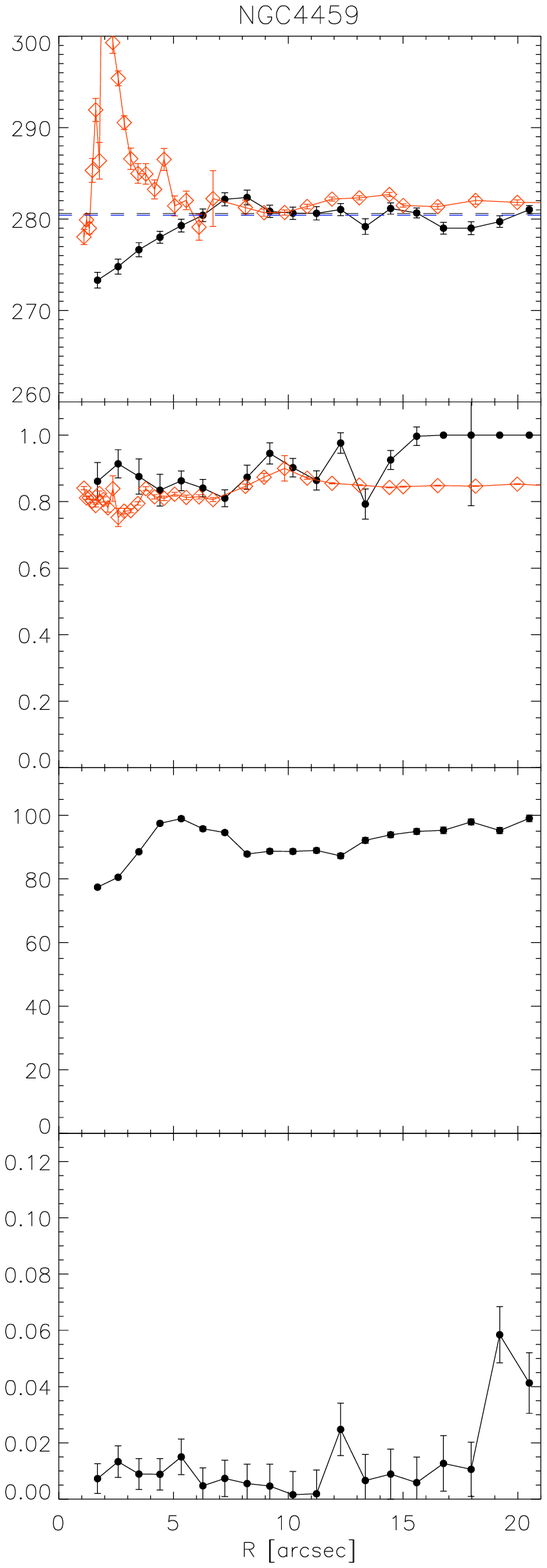} 
 \epsfxsize=0.24\textwidth \epsfbox{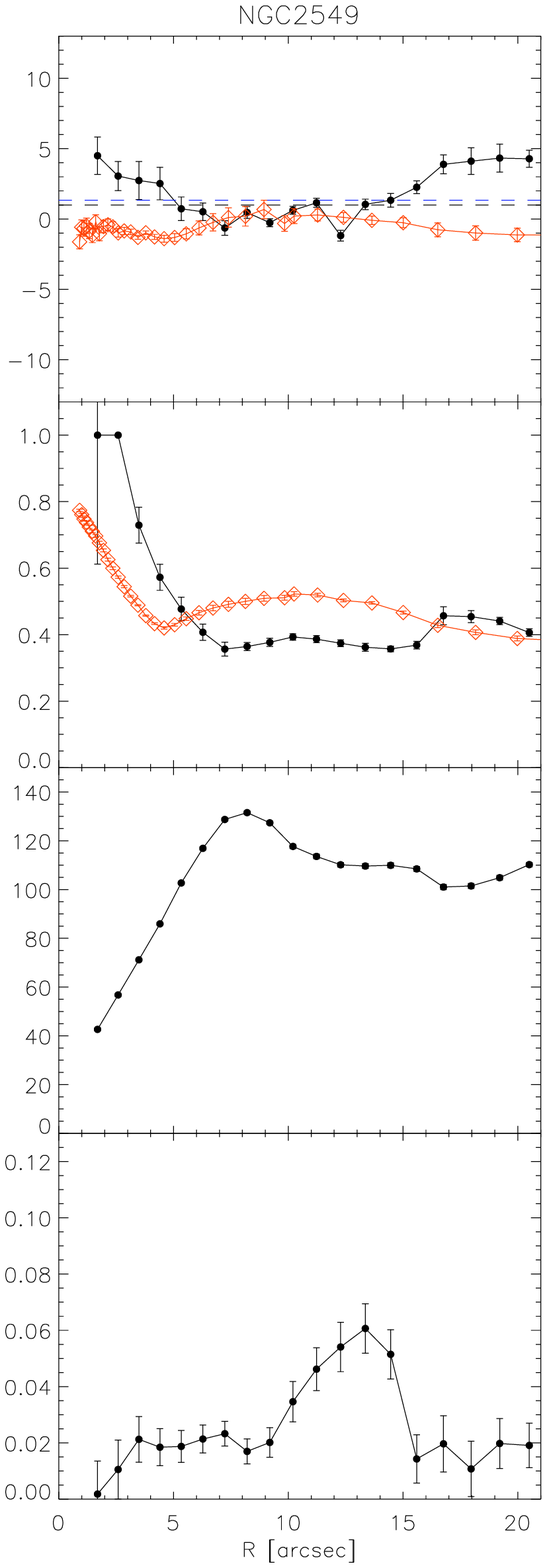} 
 \epsfxsize=0.24\textwidth \epsfbox{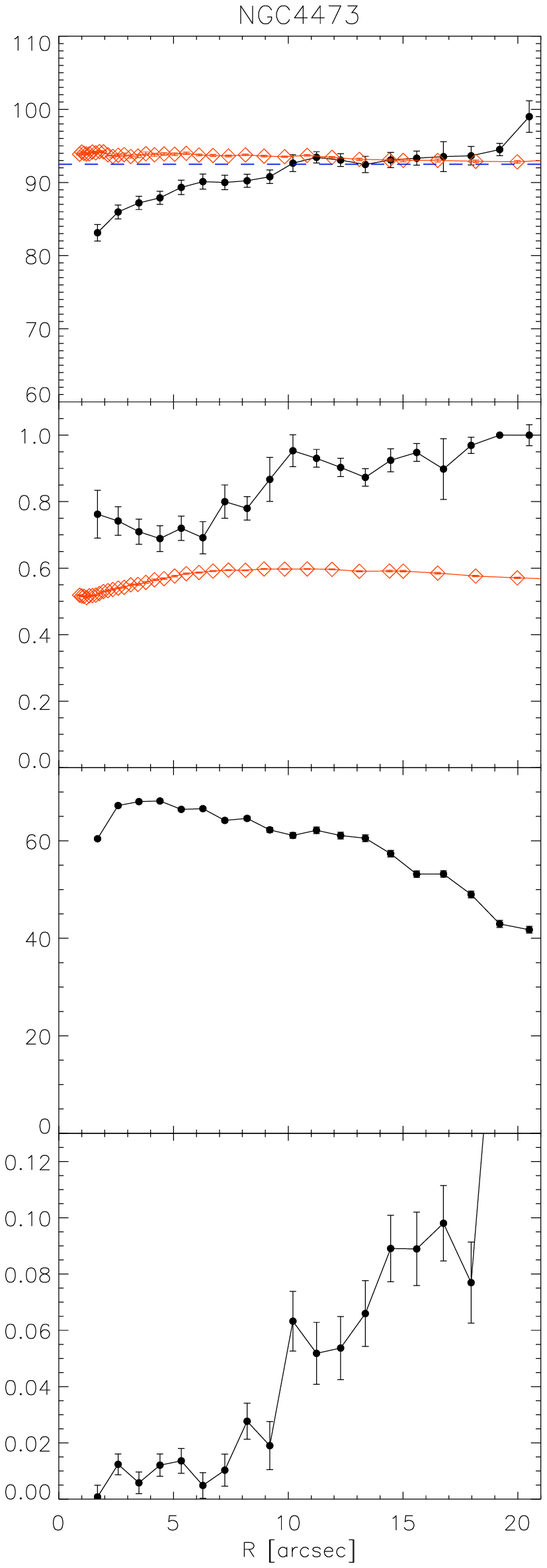}}}
\caption{\label{f:mosaic} 
  Application of the method to {\tt SAURON} observations of NGC~2974,
  NGC~4459, NGC~2549 and NGC~4473. Each panel shows kinemetric
  coefficients for the velocity maps: the kinematic position angle,
  axial ratio of the best fitting ellipses, and harmonic terms: $k_1$
  and $k_5$. Dashed horizontal lines are global kinematic position
  angle determined as the median of $\Gamma$ (blue line) and with the
  method outlined in Appendix~\ref{s:kinpa}.  They are almost always
  indistinguishable.  For comparison, the photometric values for the
  position angle and the flattening of the best fitting photometric
  ellipses are overplotted (diamonds). }
\end{figure*}
%%%%%%%%%%%%%%%%%%%%%%%%%%%%%%%%%%%%%%%%%%%%%%%%%%%%%%%%%%%%%%%%%%%%%

\noindent {\bf NGC~2974} is largely consistent with axisymmetry 
(but see \citet{2003MNRAS.345.1297E} and \citet{2005MNRAS.357.1113K}
for bar signatures), and the simplicity of the velocity map is
reflected in the kinemetric coefficients: the $k_5$ term is below
1-2\%, while the flattening and position angle are both constant over
the map except in the central few arcsecs where the departures from
the axisymmetry are the strongest. The photometric and kinematic
position angles and, especially, the flattenings agree very well.

\noindent{\bf NGC~4459} 
is a kinematically round map with large opening angles reflected in
the high values of $q$. The position angle of the velocity map changes
in the central region but remains constant beyond 7\arcsec. In the
same region, the velocity reaches a maximum and then decreases, with a
hint of a rise towards the edge of the map.  Similarly to NGC~2974,
the $k_5$ coefficient is small and below 2\% in the investigated
region.  It is possible that there is a separate central velocity
component, indicated by the change in $\Gamma$ and $k_1$. The
similarity with NGC~2974 is also reflected in the comparison with the
photometric parameters: the position angles agree very well beyond
5\arcsec\, while flattenings agree over most of the map.

\noindent {\bf NGC~2549} has a multi-component velocity map. Its central 
region ($r<5\arcsec$) is described by a linear rise in the velocity
and a low flattening, rising up to $7\arcsec-8\arcsec$ where the
velocity reaches its maximum. Further away the velocity drops and
beyond $10\arcsec$ so does the flattening (changing again at
$r>17\arcsec$).  These features are also recognisable on the velocity
map. The behaviour of the $k_5$ term is, however, not obvious from the
map. It is on average larger than in the two previous examples over
the whole map, and it shows a clear rise between 10 and 15\arcsec,
marking the change between the two components.  The kinematic and
photometric position angles differ inside 7\arcsec\, and outside
15\arcsec, while flattening of the contours is equal only beyond
15\arcsec, but a general similarity between the two flattenings is
noticeable.

\noindent {\bf NGC~4473} is our final example. The decrease in the velocity 
along the major axis coincides with the rise of the flattening and
$k_5$. The kinematic flattening does not match well the flattening of
the photometric ellipses, although the position angles agree very well
in the region of the velocity decrease. 

Fig.~\ref{f:residuals} shows the reconstructed (kinemetric velocity)
and residual (constructed as in Fig.~\ref{f:maps}) maps in the second
and third row, respectively. All velocity maps are well reconstructed
with a small number of terms. NGC~2974 and NGC~4459 have small
residuals, and the features on their maps are well described with only
$k_1$ terms. Both NGC~2549 and NGC~4473, however, have a strong $k_5$
term.  In section~\ref{ss:kinanal}, using model maps, we showed that
large $k_5$ is likely to indicate the existence of two kinematic
components. The relative orientation of the components can be any, but
we draw the attention of the reader to the co- and counter-rotating
cases presented by {\it model 0} and {\it model 180}.  Precisely the
same behaviour is visible in Figs.~\ref{f:residuals}
and~\ref{f:mosaic} for NGC~2549 and NGC~4473, respectively.

The two co-rotating velocity components are clearly visible in the map
of NGC~2549: a central fast rotating component with high flattening
and an outer slower rotating component with somewhat smaller
flattening.  On Fig.~\ref{f:residuals} it can be seen that along the
major axis the contribution of the residuals is in the same direction
as the bulk rotation, confirming the co-rotation of the components.

NGC~4473 is not as clear a case. The velocity map
(Fig.~\ref{f:residuals}) does not show any clear signature of the two
kinematic components which are inferred from the higher-order
kinemetric terms.  The residual map (Fig.~\ref{f:residuals}), however,
shows that the residuals are opposite to the bulk rotation, suggesting
that the two components are actually counter-rotating.  This
configuration can naturally explain the observed decrease of rotation
with radius.  It is thus likely that NGC~4473 is made up of two
components: a dominant one responsible for the bulk rotation of the
stars, and a lighter counter-rotating one (perhaps a disk because of
its confined contribution along the major axis) which brings more
weight to large radii. A confirmation of this result is given by
\citet{2005CQGra..22S.347C}, who constructed Schwarzschild dynamical
models of this galaxy and recovered two counter-rotating components in
the distribution function.

The velocity maps presented in this section are by no means meant to
be representative of all possible velocity structures. It is, however,
striking to see how the assumption of the method (eq.~\ref{eq:tilt})
is satisfied to a level of 2\% or better for a number of real
galaxies.  On the other hand, some galaxies do show strong deviations
in the higher-order terms, corresponding to substructure and multiple
kinematic components.  The correspondence between photometric and
kinemetric parameters is also interesting.  While it is expected that
the two position angles should agree if a galaxy is axisymmetric, it
is not expected that the flattening of the isophotes and the kinematic
ellipses will be the same.  This is satisfied in the case of thin
disks, where $q_{phot}=q_{kin} = \cos(i)$, but this relation does not
necessarily apply to spheroidal systems.  The correspondence between
$q_{phot}$ and $q_{kin}$ for spheroidal galaxies is likely related to
the degree of anisotropy of the galaxies. It may also indicate that
these galaxies contain stellar disks.

\subsection{Characterisation of features on stellar velocity maps}
\label{ss:defs}

It is possible to describe the variety of structures present in
velocity maps of galaxies, as seen in \citet{2004MNRAS.352..721E}, or
observed with other integral-field units, using the results of our
analysis.  The features can be present uniformly over a whole map or
can be different between the inner and outer parts of the map. Here we
offer a characterisation of features on the velocity maps (or the maps
themselves) based only on the kinemetric parameters. Building on the
results from the previous sections, velocity features can be sorted in
at least the following four descriptive groups:

\noindent {\bf Disk-like Rotation (DR).} This group consists of the simplest
and featureless velocity maps. A DR map shows clear rotation where the
amplitude of $k_1$ is substantial, but not necessary constant or
rising.  The $k_5$ coefficient is consistent with zero, while the
kinematic position angle $\Gamma$ and the flattening $q$ remain
constant.

\noindent {\bf Multiple Components (MC).} Similar to our {\it model 0}, MC
maps exhibit features which can be related to kinematically different,
but aligned components. The components are kinematically separate if
the flattening changes in the transition region, or if it is different
for each component. Additionally, MC are detected by a rise of the
$k_5$ coefficients in the transition region and possibly but not
necessarily by a drop of $k_1$.

\noindent {\bf Kinematic Twists (KT).}  
Twists in the iso-velocity contours are characterised by a change in
the kinematic position angle $\Gamma$. Additionally, several
characteristics of the MC group can be present as well, e.g. $k_5 > 0$
or drop in $k_1$, indicating different kinematic components.

\noindent {\bf Low-level Rotation (LR).} 
A rather special feature on the velocity maps occurs when there is no
detectable rotation. This effect is dependent on the instrumental
resolution (rotation is too low to be measured), but its origin is in
the formation scenario and intrinsic distribution of orbits. This
group can be characterised by a requirement that $k_1$ remains below a
certain limiting value of $k_1'$. Maps belonging to this class are
characterised by high amplitude of $k_5$, often bigger then $k_1$.
Due to the indeterminacy of $q$ and $\Gamma$ for the LR type, the
analysis should be performed along circles, which changes the meaning
of the harmonic terms.

It is important to note that the exact details, such as the limiting
value of $k_1'$, or the tolerated change in $\Gamma$, which serve to
characterise the structures, depend on the instrumental settings and
the quality of the observed data. However, it is clear that the
kinemetric parameters could be used to describe and classify the
properties of velocity maps.

%%%%%%%%%%%%%%%%%%%%%%%%%%%%%%%%%%%%%%%%%%%%%%%%%%%%%%%%%%%%%%%%%%
%
%SECTION6 SECTION6 SECTION6 SECTION6 SECTION6 SECTION6 SECTION6 
%
%%%%%%%%%%%%%%%%%%%%%%%%%%%%%%%%%%%%%%%%%%%%%%%%%%%%%%%%%%%%%%%%%%

\section{Summary}
\label{s:sum}

We have presented a generalisation of surface photometry to the
higher-order moments of the LOSVD, observed with integral-field
spectrograph. We call our method \emph{kinemetry}. For even moments of
the LOSVD, kinemetry reduces to photometry. For odd moments, kinemetry
is based on the assumption that it is possible to define an ellipse
such that the kinematic profile extracted along the ellipse can be
well described by a simple cosine law. This assumption is satisfied in
the case of simple axisymmetric rotators.  Kinematic profiles that
deviate from axisymmetry or contain multiple kinematic components will
be more complex, and the residuals can be measured through a harmonic
analysis of the profiles. We also find that the assumption is well
satisfied for a number of observed galaxies.

Since the profiles are extracted along the best fitting ellipse, the
number of harmonic parameters necessary to describe the profile is
usually small.  A total of six harmonic terms is generally necessary
to determine the best fitting ellipse ($A_0$, $A_1$, $A_2$, $B_2$,
$A_3$ and $B_3$). The deviations from the simple cosine law are
carried in higher harmonic terms. It is not expected that many more
higher-order terms will be necessary in the general case, and we stop
the analysis at the $A_5$ and $B_5$ harmonics.

The most important parameters of the analysis are the kinematic
position angle $\Gamma$, flattening $q$ of the ellipses, and the
harmonic terms: $k_1$ and $k_5$, where we combined the terms of the
same order.  We briefly summarise their properties:

\begin{itemize}
  
\item[-] the kinematic position angle traces the maximum velocity and
  describes the orientation of the map.
  
\item[-] the flattening is related to the opening angle of the
  iso-velocities and describes the different components of the maps,
  which can appear {\it flat} or {\it round}. In the special case of a
  thin disk it gives the inclination of the galaxy;
  
\item[-] $k_1$ is the dominant harmonic coefficient and describes the
  amplitude of bulk motions. It is related to the circular velocity.
  Higher order coefficients, which we normalise to $k_1$, describe the
  deviations from simple circular motion;
  
\item[-] $k_5$ is the first higher-order term which is not fitted and
  quantifies higher-order deviations from rotational motions. It
  indicates complexity on the maps and is sensitive to the existence
  of separate kinematic components. A photometric analog of this term
  is the coefficient that describes the diskiness/boxiness of the
  isophotes.
\end{itemize}

This method can be straight-forwardly applied to all moments of the
LOSVD observed from different objects (e.g. early- and late-type
galaxies). However, the analysis and the intepretation of the harmonic
terms has to be related to the physical nature of the observed
objects. An example are velocity maps of thin disks. In this case, the
kinemetric flattening and position angle are directly related to the
inclination and orientation of the disk, and the results of kinemetry
are equal to results of the tilted-ring method. If $q$ and $\Gamma$
are kept fixed, kinemetry reduces to the harmonic analysis of
\citet{1997MNRAS.292..349S} and \citet{2004ApJ...605..183W}. In this
case, since the velocity profiles are not extracted along the best
fitting ellipses, all harmonic terms will carry information about
departures from the simple circular motion.

We applied the method on one model and four observed galaxies. The
model galaxy was a two-integral model of an observed axisymetric
galaxy. We constructed maps of observable odd kinematic moments of the
LOSVD: mean velocity, velocity dispersion and Gauss-Hermite
coefficient $h_3$. We analysed both maps to show that the method works
well on an axisymmetric galaxy (all higher-order terms were small or
consistent with zero). We also introduced an intrinsic scatter and
currently realistic measurement uncertainties to show the difficulties
expected analysing the higher-order odd moments (e.g. $h_3$) obtained
from state-of-the-art integral-field observations. The position angle,
flattening and amplitude of the moment can be realistically recovered,
but the signal-to-noise ratio is too low for expansion to higher-order
harmonics.

The four galaxies were taken from the {\tt SAURON} sample. We used them
to show the descriptive power of the new method: detection of co- or
counter-rotating subcomponents well hidden in the kinematics, which
could be detected previously only through detailed modelling. We also
showed that there are early-type galaxies with rotation velocity
described by a simple cosine law. The kinemetric position angle and
flattening of these galaxies are in a good agreement with the
photometric position angle and flattening. 

Based on the kinemetric parameters we characterise several features on
the velocity maps: Disk-like Rotation, Multiple Components, Kinematic
Twists and Low-level Rotation.  Other, more specific and instrument
dependant features on the maps can be constructed from these basic
groups.

We wish to stress that the method works well on velocity maps because
many of them satisfy the basic assumption of the method to the level
of a few per cent. This empirical fact is related to the internal
structure of the galaxies. We developed the method of kinemetry to
harvest the information from the observed maps of moments of the LOSVD
probing the nature of galaxies.

An IDL implementation of the described algorithm is available on the
web address: http://www-astro.physics.ox.ac.uk/$\sim$dxk/.\\

\noindent{\bf Acknowledgements}
We thank Eric Emsellem for providing a model axisymmetric galaxy and
Maaike Damen for providing the photometric data used in this work. DK
thanks Martin Bureau and Eric Emsellem for a careful reading of the
manuscript and Marc Sarzi, Glenn van den Ven and Richard McDermid for
fruitful discussions. This research was supported by NOVA, the
Netherlands Research School for Astronomy and by PPARC grant
PPA/G/O/2003/00020 'Observational Astrophysics in Oxford'. MC
acknowledges support from a VENI grant 639.041.203 awarded by the
Netherlands Organization for Scientific Research (NWO).

%%%%%%%%%%%%%%%
% Reference List
%%%%%%%%%%%%%%%   

%\bibliographystyle{mn2e} \bibliography{../refs.bib}

%%%%%%%%%%%%%%%
% Appendices
%%%%%%%%%%%%%%%

%\clearpage
\appendix
\section{Extraction along circles}
\label{s:circ}

The simplest case of kinemetric expansion is by extracting velocity
profiles along concentric circles. This approach has several
advantages over the more general case of 'ellipse fitting'. They are:

\begin{itemize}
\item[-] there is no a priori assumption about the structure of the map
  
\item[-] extraction of harmonic coefficients is a purely linear problem
  (no degeneracy between parameters)

\item[-] reconstruction of maps is straight-forward even in the most
  complicated cases
\end{itemize}

These points, however, are weighted against the fact that the number
of the extracted harmonic terms which describe a more complicated map
is large, and often the coefficients cannot be clearly associated with
the physical properties of the objects. In this Appendix, for
completeness, we briefly summarise the analysis of
\citet{2001sf2a.conf..289C} and \citet{2004KrajnovicThesis}.

Requiring that axial ratio $q = 1$, kinemetry reduces to Fourier
expansion along concentric circles via eq.~(\ref{eq:kinlin}), where
the coefficients $A_{n}, B_{n}$ are determined by a least-squares fit
with a basis $\{1, \cos\theta, \sin\theta,...,\cos N\theta, \sin
N\theta\}$, where $\theta$ is now the polar angle, to which the
eccentric anomaly, $\psi$, reduces in this case. Depending whether the
fit is to an odd or an even kinematic moment, the terms in the
expansion can be selected such that $n$ is odd or even, respectively.
In a more compact way, the harmonic series is presented by the
eq.~(\ref{eq:kinemetry}), where the amplitude and phase coefficients
($k_{n}$, $\phi_{n}$) are easily calculated from the $A_{n}, B_{n}$
coefficients following eq.~(\ref{eq:coeff}).

The following description of the harmonic coefficients is similar to
the description of the coefficients obtained by ellipse fitting, but
differs in some crucial details.

The zeroth-order term, $A_{0}$, measures the mean level of the map.
For the first kinematic moment, this is equivalent to the systemic
velocity of the galaxy.  For $h_{3}$ maps, which have mean level equal
to zero by construction, $A_{0}$ will, like other even terms, be
consistent with zero.

The coefficient $k_{1}$ gives the general shape and amplitude of the
odd moment map and is the dominant term. The first-order correction is
given by the next odd term, $k_{3}$. This term can be named the
\emph{morphology} term because it describes most of the additional
geometry of the map. Often, the first two odd terms are enough to
describe the velocity map, although the kinemetric expansion may
require higher terms as well, $k_{5}$ and even $k_{7}$.

Connected to the amplitude terms are the corresponding phase terms,
$\phi_{n}$. They determine the orientation of the map, but their
contribution depends on the relative strength of the amplitude terms.
The first phase coefficient, $\phi_{1}$, gives the mean position angle
of the velocity map (measured from the horizontal axis, $\theta =0$).
The angle that $\phi_{1}$ measures is the position angle of the
$k_{1}$ term. This is, in general, slightly different from the
positions of the maxima on the map, which are also influenced by the
contribution from higher-order terms. However, it does give the global
orientation of the map.  This phase corresponds to previously defined
kinematic position angle $\Gamma$.

The angle $\phi_{3}$ is the phase angle of the third harmonic: the
next significant term. For a small amplitude of $k_{3}$, its
contribution to the overall orientation will be small, and the
position of the maximum velocity will be given accurately by
$\theta_{max} = \phi_{1}$. The phases of higher-order terms are
interesting because, for an axisymmetric galaxy, they satisfy the
relation:

\begin{equation}
  \label{eq:axi_phase}
  \phi_{1} - \phi_{i} = \frac{n\pi}{i}
\end{equation}

\noindent where $n \in \mathbb{Z}$ and $\phi_i$ is the $i$-th order
term in the expansion.

This condition can be easily derived considering that for an
axisymmetric velocity map the position of the zero velocity curve, the
curve along which the velocity is zero on the map, is orthogonal to
the kinematic angle given by $\phi_1$. This means that K$(a,\theta)$ =
0, for $\theta=\phi_{1} + \pi/2$. Neglecting the higher order terms,
($k_{i} > k_{3}$), and substituting $\theta=\phi_{1} + \pi/2$ into
eq.~(\ref{eq:kinemetry}) one obtains the result of
eq.~(\ref{eq:axi_phase}).

Alternatively, deviations from axisymmetry can be quantified following
eqs.~(\ref{eq:kinemetry}) and~(\ref{eq:axi_phase}). If K$(a,\theta
=\phi_{1} + \pi/2) = \Delta V \ne 0$, then one finds:
\begin{equation}
  \label{eq:non_axi}
  \frac{\Delta V}{k_{1}} = \frac{k_{3}}{k_{1}} \sin3(\phi_{1} -
  \phi_{3}),
\end{equation}

\noindent where we express the relation as a ratio of the dominant
term in the expansion. This relation quantifies the contribution of
the $k_3$ term due to departures from axisymmetry and can be
generalised to other higher terms. In cases of large misalignments
between the kinematic and photometric position angles, $\phi_{1}$
should be replaced by the photometric PA in the above equation.

In many ways, analysis of the even moments is analogous to
that for the odd moments, taking the proper parity into account.

The dominant term of the even maps is $A_{0}$, and it describes the
absolute level of the map as a function of radius. The next important
term in the expansion is $k_{2}$, and it is the \emph{morphology} term
of the even moment maps, which describes features such as elongation
along the minor axis or ``bow-tie'' shapes, often seen in observed
velocity dispersion maps \citep{2004MNRAS.352..721E}. The orientation
of the morphologically distinct features are determined by the phase
angles $\phi_{2}$ of the $k_{2}$ term.

The parity of kinematic moments, expressed by eqs.~(\ref{eq:paspolar})
and (\ref{eq:maspolar}) in section~\ref{s:back}, has the following
consequences on the harmonic expansion along circles:
\begin{itemize}
\item [-] point-symmetry requires only even terms in the harmonic
  expansion
\item [-] point-anti-symmetry requires only odd terms in the harmonic
  expansion
\item [-] additional mirror-(anti)-symmetry requires only cosine terms
  in the harmonic expansion
\end{itemize}

It follows that kinemetry can easily be used for constructing
two-dimensional kinematic maps of a given symmetry. This is done by
fixing certain harmonic terms (odd, even or sine terms) to zero; e.g.
a bi-anti-symmetric map is created by keeping only odd cosine terms in
the harmonic expansion.  Additionally, if the number of harmonic terms
in the expansion is small, the reconstructed map will be smoothed. In
this way, kinemetry can be used for removing the higher-order
harmonics from the data and effectively filtering the map.

\section{Influence of incorrect centre, flattening and orientation}
\label{s:influ}

%%%%%% Figure ap1%%%%%%%%%%%%%%%%%%%%%%%%%%%%%%%%%%%%%%%%%%%%%%%%%%%%
\begin{figure}
  \includegraphics[width=\columnwidth]{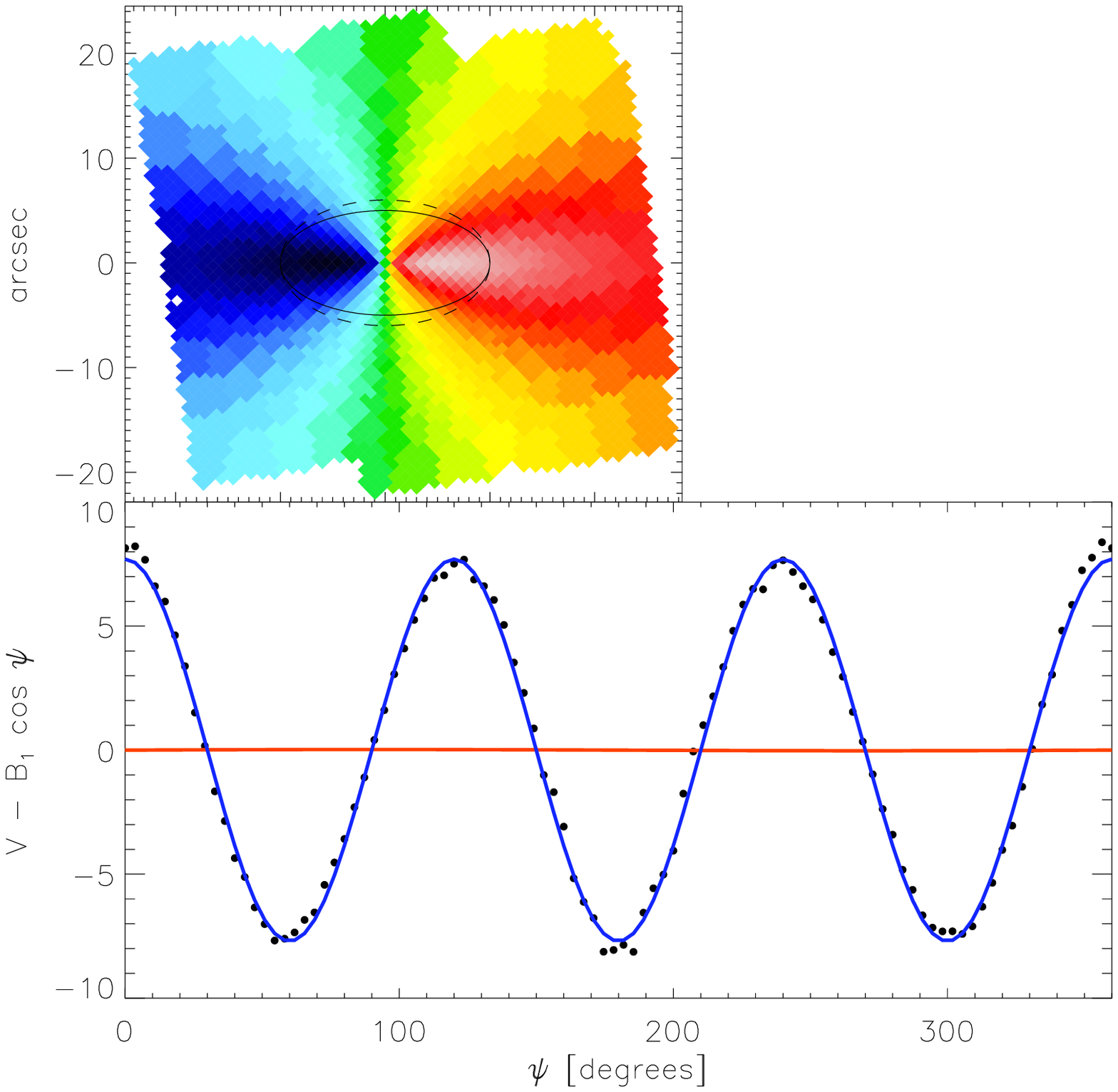}
  \caption{\label{f:inc} 
    The effect of an incorrect flattening on the coefficients of
    harmonic expansion of a velocity profile. Upper panels shows the
    model velocity map having constant orientation (kinematic position
    angle) and flattening.  The solid ellipse has correct flattening
    and orientation, while the dashed ellipse has incorrect axial
    ratio and correct orientation. The lower panel shows residuals
    between the extracted velocity profile and the circular velocity
    $B_1 \cos(\psi)$ (solid symbols). Overplotted are contributions
    of $A_1 \sin(\psi)$ (red line) and $ B_3 \cos(3 \psi)$ (blue
    curve). }
\end{figure}
%%%%%%%%%%%%%%%%%%%%%%%%%%%%%%%%%%%%%%%%%%%%%%%%%%%%%%%%%%%%%%%%%%%%%%
%%%%%% Figure ap2%%%%%%%%%%%%%%%%%%%%%%%%%%%%%%%%%%%%%%%%%%%%%%%%%%%%
\begin{figure}
        \includegraphics[width=\columnwidth]{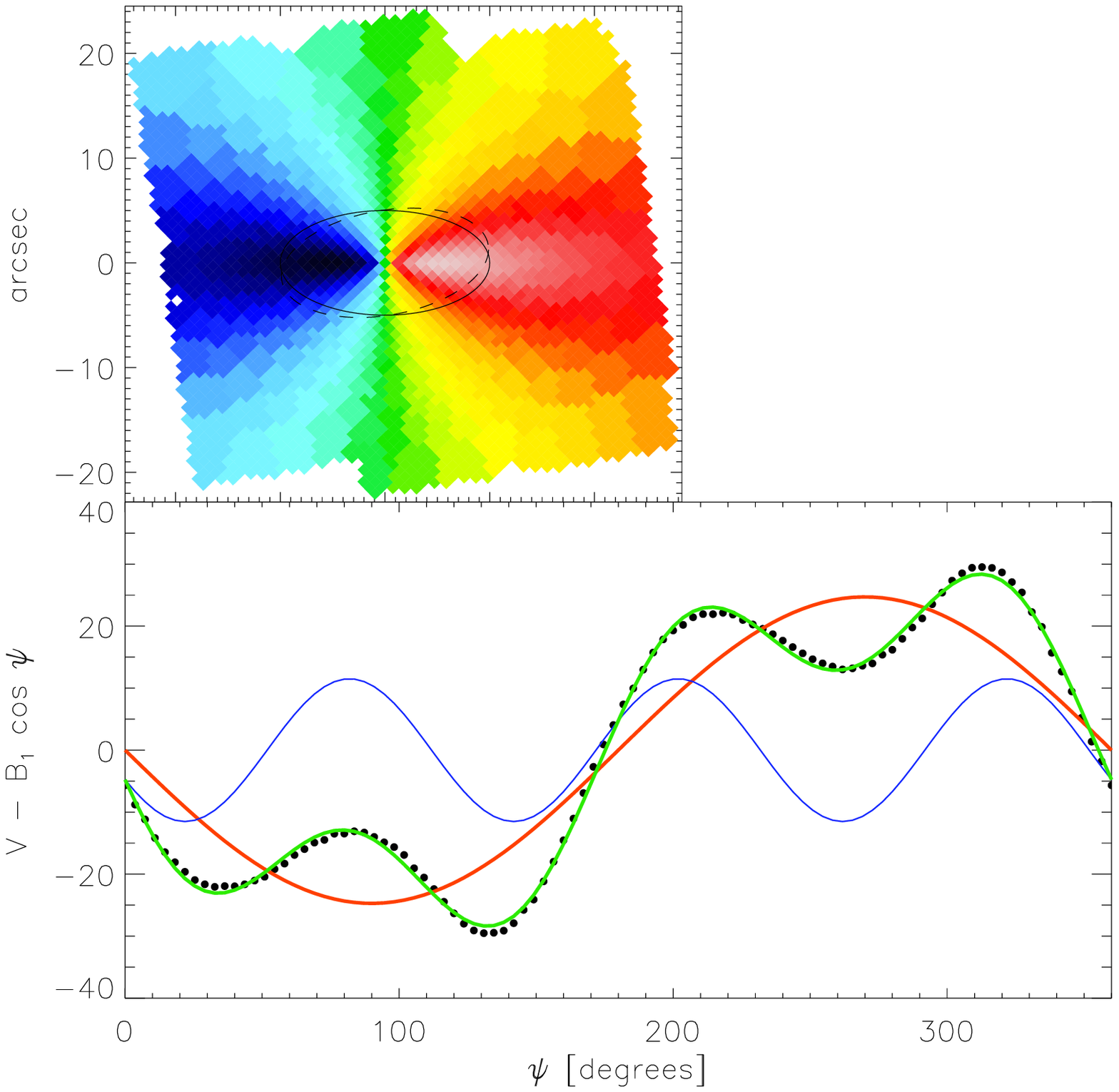}
  \caption{\label{f:kpa} 
    Same as Fig.~\ref{f:inc} where the dashed ellipse has correct
    flattening but incorrect orientation.  Next to the residuals on
    the lower panel overplotted are curves indicating the contribution
    of different terms: $A_1 \sin(\psi)$ (red curve), $A_3
    \sin(3\psi) + B_3 \cos(3 \psi)$ (blue curve), and a
    combination $A_1 \sin(\psi) + A_3 \sin(3\psi) + B_3 \cos(3
    \psi)$ (green curve). }
\end{figure}
%%%%%%%%%%%%%%%%%%%%%%%%%%%%%%%%%%%%%%%%%%%%%%%%%%%%%%%%%%%%%%%%%%%%%%
%%%%%% Figure ap3%%%%%%%%%%%%%%%%%%%%%%%%%%%%%%%%%%%%%%%%%%%%%%%%%%%%
\begin{figure}
        \includegraphics[width=\columnwidth]{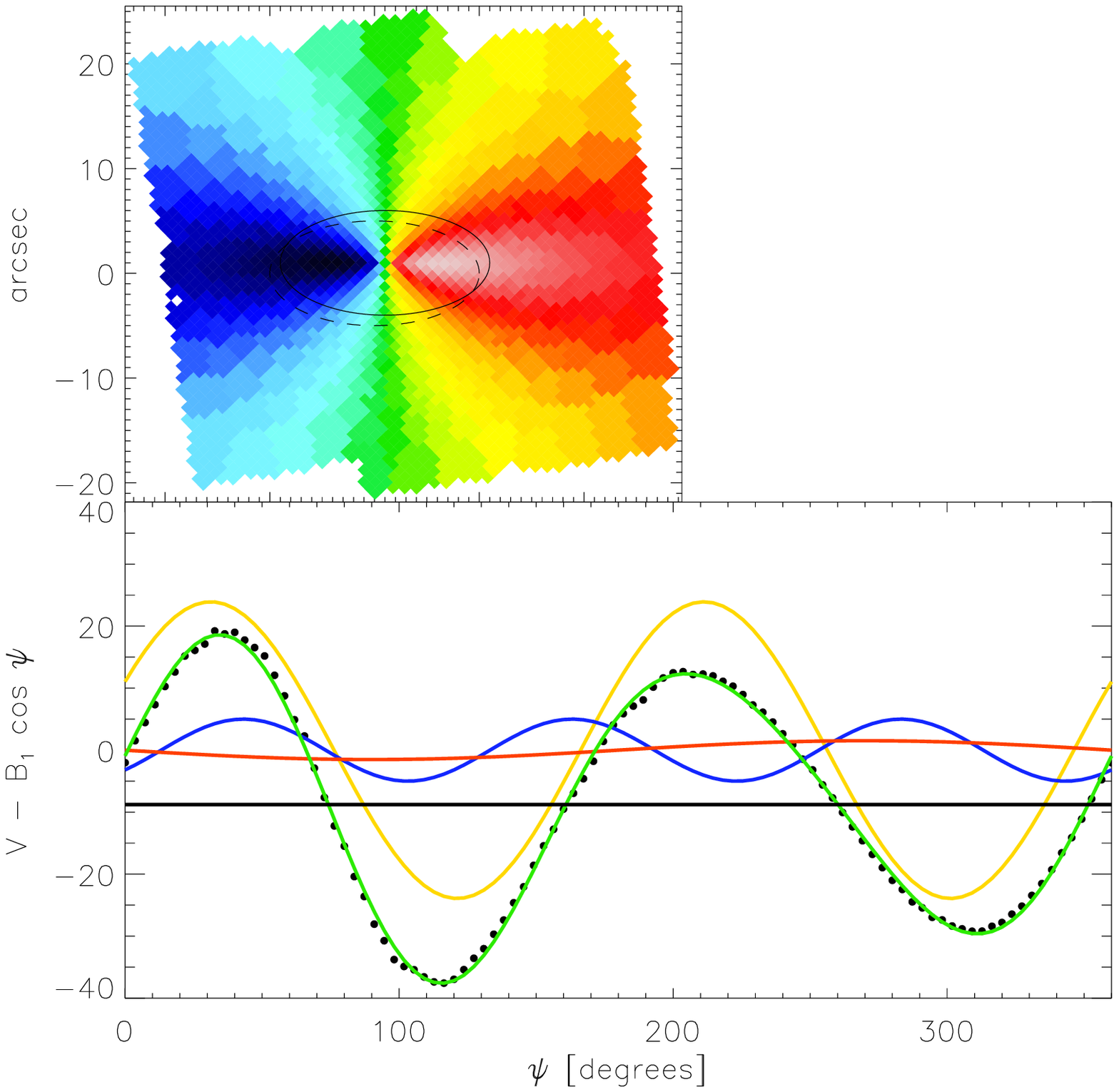}
  \caption{\label{f:cen} 
    Same as Fig.~\ref{f:inc} where the dashed ellipse of correct
    flattening and position angle was shifted by 1\arcsec horizontally
    and 1\arcsec vertically from the centre. The lower panel presents
    the residual and contributions from different terms: $A_0$ (black
    line), $A_1 \sin\psi$ (red line), $A_2 \sin(2\psi) + B_2
    \cos(2\psi)$ (orange curve) and $A_3 \sin(3\psi) + B_3
    \cos(3\psi)$ (blue curve). The combined contribution of all
    these terms is overplotted with the green curve. }
\end{figure}
%%%%%%%%%%%%%%%%%%%%%%%%%%%%%%%%%%%%%%%%%%%%%%%%%%%%%%%%%%%%%%%%%%%%%%

Here we test the sensitivity of harmonic expansion of a velocity
profile given by eq.~(\ref{eq:kinNew}) with respect to a chosen
incorrect flattening (inclination in a thin disk case), position angle
and the centre of the ellipse along which the profile was extracted.
We constructed a simple velocity map using the circular velocity form
the Hernquist potential and an exponential light profile corresponding
to the central (disk) component of the {\it model 0}. The velocity map
had a constant flattening of $q_0=0.5$ and kinematic position angle of
$\Gamma_0=0\degr$. Systemic velocity was set to zero. We first
extracted a velocity profile from this map at $a=10\arcsec$ along a
pre-designed ellipse. The ellipse was first chosen to have an
incorrect flattening $q' = 0.6$, but correct kinematic position angle
$\Gamma_0$ (Fig.~\ref{f:inc}). We repeated the exercise selecting the
ellipse with incorrect $\Gamma' = 10\degr$ and correct $q$
(Fig.~\ref{f:kpa}).  Each of the profiles were then harmonically
expanded with eq.~(\ref{eq:kinNew}). Finally, we shifted an ellipse,
defined by correct values of $q$ and $\Gamma$, horizontally and
vertically for 1\arcsec (Fig.~\ref{f:cen}). The corresponding velocity
profile was expanded using first $N=6$ (even and odd) terms in
eq.~(\ref{eq:kinlin}).

Harmonic expansion of a profile extracted along an ellipse defined by
the correct centre, flattening and orientation would (by
construction) yields all harmonic coefficients zero expect the $B_1$,
which gives the circular velocity of the map. Hence, on the lower
panels of Figs.~\ref{f:inc}, \ref{f:kpa} and \ref{f:cen} we plot the
difference (residuals) of the extracted velocity profile and the $B_1
\cos(\psi)$ term.

On Fig.~\ref{f:inc} overplotted are the contributions of terms: $A_1
\sin(\psi)$ and $B_3 \cos(3 \psi)$. The $A_3 \sin(3 \psi)$ is
nearly zero, like the $a_1$, and is not shown. The contribution of
only the $B_3 \cos(3 \psi)$ term can explain the distribution of
residuals, confirming that only $B_3$ coefficient is affected when
choosing an incorrect flattening.

On Fig.~\ref{f:kpa} overplotted are the following terms: $A_1
\sin(\psi)$, $A_3 \sin(3\psi) + B_3 \cos(3 \psi)$, $ B_3 \cos(3
\psi)$, as well as the combination $A_1 \sin(\psi) + A_3
\sin(3\psi) + B_3 \cos(3 \psi)$. All of these terms contribute
significantly to describe the distribution of residuals. The dominant
coefficient is $A_1$, while $A_3$ and $B_3$ contribute in this example
with about 45\% and 15\% in amplitude, respectively. In the limit of a
small incorrect position angle, \citet{1997MNRAS.292..349S} showed
that only $A_1$ and $A_3$ are affected, but our example shows that for
larger incorrect position angles the $B_3$ term has to be considered
as well.

Fig.~\ref{f:cen} presents contributions for $A_0$, $A_1$, $A_2$,
$B_2$, $A_3$ and $B_3$ harmonic terms. To the first order, only $A_0$,
$A_2$ and $B_2$ terms are sensitive to incorrect centring. The same
result was also analytically derived by \citet{1997MNRAS.292..349S}, in
the limit of small miscentering. However, for large deviation from the
true centre coordinates, contribution from the $A_3$ and $B_3$ terms
becomes important, while $A_1$ remains much smaller. 

We conclude this exercise by stating that the ellipse parameters for
an odd kinematic moment map can determined by harmonic expansion of
the $A_0$, $A_1$, $A_2$, $B_2$, $A_3$ and $B_3$ terms only.

\section{Determination of global kinematic position angle}
\label{s:kinpa}

The apparent angular momentum {\bf L${_p}$} of a galaxy, as a
projection of the intrinsic angular momentum to the plane of sky, is
defined as \citep{1988FranxThesis}:

\begin{equation}
  \label{eq:angular}
 \mathbf{L_p} = \int_{P} \mathbf{r_p} \times \Sigma \mathbf{v_r} d\mathbf{r_p}
\end{equation}

\noindent
where $\mathbf{r{_p}}$ is a projection to the plane of sky of the
vector $\mathbf{r}$ at which the mean velocity of stars, $\mathbf{v}$,
projects to the observed radial velocity vector $\mathbf{v{_r}}$.
$\Sigma$ is surface density and the integration is over the whole
projection plane. As noted by \citet{1991ApJ...383..112F}, it follows
from eq.~(\ref{eq:angular}) that the apparent angular momentum is
fully specified by the observed surface brightness and the observed
two dimensional velocity map. In addition to that, if figure rotation
is absent, the apparent angular momentum is parallel to the projection
of the intrinsic angular momentum. This means that the projection of
the full three-dimensional velocity space is reduced to the projection
of the angular momentum.  Reversing the argument, measuring the
(luminosity weighted) kinematic position angle of the two-dimensional
projection of velocities, it is possible to determine the position of
the apparent angular momentum.

The position angle of the apparent angular momentum $\Gamma_{kin}$ can
be determined from eq.~(\ref{eq:angular}) using integral-field
observations of galaxies. This is the most physical way of
determination of $\Gamma_{kin}$, but it is somewhat dependant on the
geometry of the maps\footnote{We do not discuss here the influence of
  the size of the maps. Clearly, only $\Gamma_{kin}$ of the observed
  part of the galaxy can be determined, because eq.~(\ref{eq:angular})
  is defined over the whole projection plane}, like the relative
orientation of the map with respect to the orientation of the apparent
angular momentum.

A more robust way of determining $\Gamma_{kin}$ is based on a direct
determination of the kinematic position angle $\Gamma$. Kinemetry
measures $\Gamma$ as a function of radius, but we are interested in a
global (independent of radius) value of the kinematic position angle,
$\Gamma_G$, because of a direct correspondence between $\Gamma_G$ and
$\Gamma_{kin}$. $\Gamma_G$ can be determined in a robust way using the
information in the whole velocity map. For any chosen $\Gamma_G$, we
construct a bi-anti-symmetric velocity map $V'(x,y)$ with the $x$-axis
along the given $\Gamma_G$. This is done by replacing the measured
mean velocity $V(x,y)$ inside each Voronoi bin, of centroidal
coordinates $(x,y)$, with the weighted average of the corresponding
velocity in the four quadrants, using linear interpolation when needed
to estimate the velocity:

\begin{equation}
  \label{eq:velo}
  V'(x,y) = \frac{V(x,y) + V(x,-y) - V(-x,y) - V(-x,-y)}{4}
\end{equation}

\begin{figure}
        \includegraphics[width=\columnwidth]{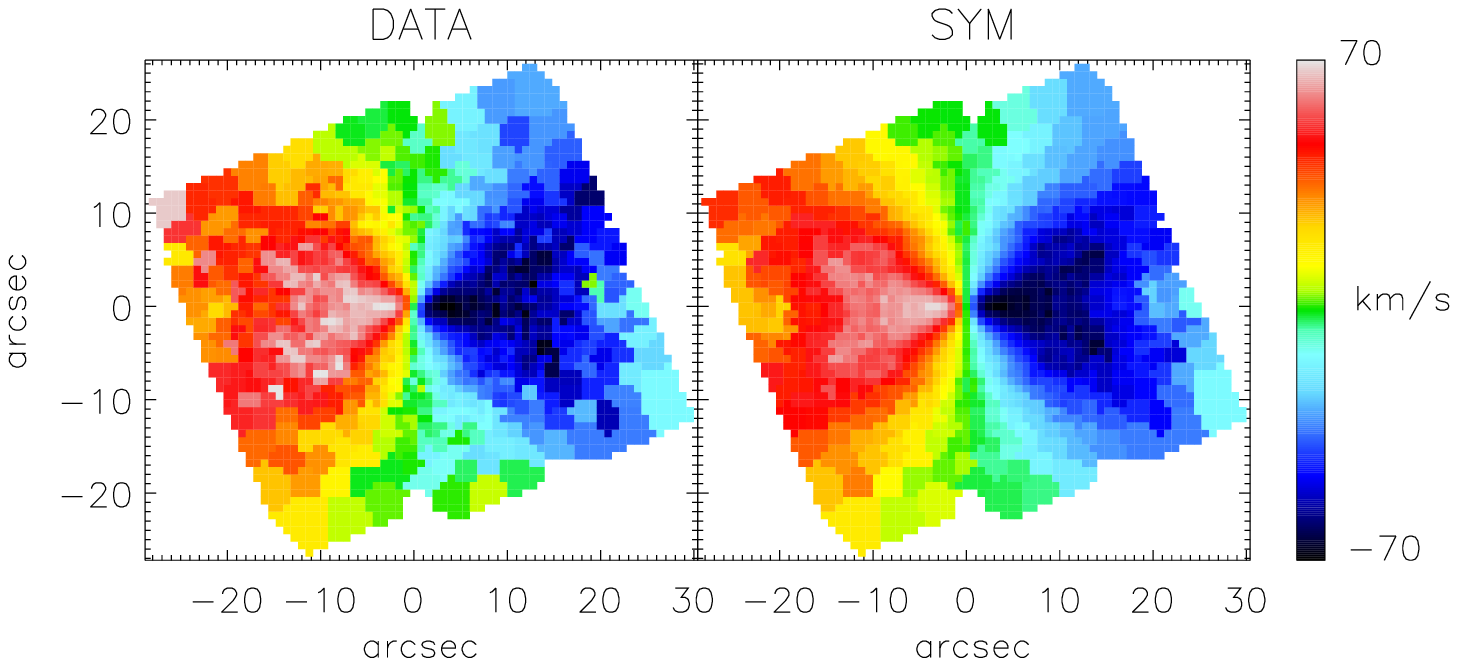}
  \caption{\label{f:sym} 
    Comparison of observed velocity map (left) and bi-anti-symmetric
    map (right) at the best fitting $\Gamma_G$ using the method
    outlined in the text. We used {\tt SAURON} observations of
    NGC~4473. North is up and east to the left. }
\end{figure}

\noindent 
When any of the four symmetric coordinates was outside the convex hull
defined by the bins centroids, we used only the remaining values to
determined $V'$. Fig.~\ref{f:sym} shows an example of a
bi-anti-symmetric map. The true $\Gamma_G$ was defined as the angle
which minimises

\begin{equation}
  \label{eq:chi2}
  \chi^2 = \sum_{n=1}^{N} \left( \frac{V'(x,y) - V(x,y)}{\Delta V(x,y)}  \right)^2
\end{equation}

\noindent
In this way it is possible to assign simple error estimates to the
obtained $\Gamma_G$ as the range of angles for which $\Delta \chi^2 <
9$, which corresponds to the $3\sigma$ confidence level for one
parameter.

We used velocity maps from section~\ref{ss:ngc4473} to compute the
global kinematic position angle. Obtained values for NGC~2974,
NGC~4459, NGC~2549 and NGC~4473 are 43.2, 280.6, 1.0, 92.5,
respectively. The results are overplotted on Fig.~\ref{f:mosaic} where
they can be compared with the kinemetric position angle, $\Gamma$,
which changes with radius describing specific features on the maps.
They are in a very good agreement as can be seen comparing with the
median values of the kinemetric position angle.

%%%%%% Figure ap4%%%%%%%%%%%%%%%%%%%%%%%%%%%%%%%%%%%%%%%%%%%%%%%%%%%%
%%%%%%%%%%%%%%%%%%%%%%%%%%%%%%%%%%%%%%%%%%%%%%%%%%%%%%%%%%%%%%%%%%%%%%

\label{lastpage}

\end{document}